\documentclass[
prd,nofootinbib,amsfonts,singlecolumn,superscriptaddress]{revtex4-1}
\usepackage{gensymb}
\usepackage{ae} 
\usepackage{aecompl}
\usepackage{bm} 
\usepackage[usenames,dvipsnames]{color}
\usepackage[pdfa,bookmarks=true,bookmarksnumbered=true]{hyperref}
 \hypersetup{
     colorlinks=true,
     linkcolor=blue,          % color of internal links (change box color 
     citecolor=Red,
     urlcolor=Mahogany
     }

\usepackage{amsmath}
\usepackage{amssymb}
\usepackage{epsfig}

\usepackage[utf8]{inputenc}

\usepackage{fontenc}
\usepackage{graphicx}
\usepackage[lofdepth,lotdepth,caption=false]{subfig}
\usepackage{color}
\usepackage{booktabs}
\usepackage{tabularx}
\usepackage{multirow}
\usepackage{longtable}
\usepackage{dcolumn}
\usepackage{rotating}%

\def\hlpm#1{\textcolor{black}{\textrm{#1}}}

\newcommand{\bi}{\begin{itemize}}
\newcommand{\ei}{\end{itemize}}

\def\beq{\begin{equation}}
\def\eeq{\end{equation}}

\newcommand{\bea}{\begin{eqnarray}}
\newcommand{\eea}{\end{eqnarray}}

\newcommand{\nue}{%\overset{(-)}
{\nu_{\mu}} \rightarrow 
{\footnotesize{\nu_{e}}}}
\newcommand{\nuebar}{\bar{\nu}_{\mu} \rightarrow \bar{\nu}_{e}}
\newcommand{\numu}{
{\nu_{\mu}} \rightarrow %\overset{(-)}
{\nu_{\mu}} 
}
\newcommand{\numubar}{\bar{\nu}_{\mu} \rightarrow \bar{\nu}_{\mu}}
\newcommand{\nutau}{%\overset{(-)}
{\nu_{\mu}} \rightarrow %\overset{(-)}
{\nu_{\tau}}}
\newcommand{\nutaubar}{\bar{\nu}_{\mu} \rightarrow \bar{\nu}_{\tau}}
\newcommand{\len}{\mathcal{R}_{\text{\tiny{LE}}}}
\newcommand{\men}{\mathcal{R}_{\text{\tiny{ME}}}}
\newcommand{\lenbar}{\mathcal{\overline{R}}_{\text{\tiny{LE}}}}
\newcommand{\menbar}{\mathcal{\overline{R}}_{\text{\tiny{ME}}}}
\newcommand{\area}[1]{\ensuremath{R_{\text{\tiny{#1}}}^{\text{area}}}}
\newcommand{\fraction}[1]{\ensuremath{R_{\text{\tiny{#1}}}^{\text{fraction}}}}
\newcommand{\ta}{\theta_{12}}
\newcommand{\tb}{\theta_{13}}
\newcommand{\tc}{\theta_{23}}
\newcommand{\ldm}{\Delta m_{31}^2}
\newcommand{\sdm}{\Delta m_{21}^2}

\newcommand{\chisq}{\Delta \chi^{2}}

\newcommand{\ie}{{\it i.e.}}
\newcommand{\eg}{{\it e.g.}}

\def\epsilon{\varepsilon}

\def\<{\langle}
\def\>{\rangle}

\def\lsim{\mathrel{\rlap{\lower4pt\hbox{\hskip1pt$\sim$}}
    \raise1pt\hbox{$<$}}}         %less than or approx. symbol
\def\gsim{\mathrel{\rlap{\lower4pt\hbox{\hskip1pt$\sim$}}
    \raise1pt\hbox{$>$}}}         %greater than or approx. symbol

\begin{document}  

\title{
Impact of high energy beam tunes on the sensitivities to the standard unknowns at DUNE
%Role of high energy beam tunes in optimizing the sensitivity to current unknowns  at DUNE
}

     \author{Jogesh Rout}
     \email{jogesh.rout1@gmail.com}
\affiliation{School of Physical Sciences, Jawaharlal Nehru University, 
      New Delhi 110067, India}  
\author{Samiran Roy}
\email{samiran@prl.res.in}
\affiliation{Physical Research Laboratory, Navrangpura,
Ahmedabad - 380 009, India}
\affiliation{Harish-Chandra Research Institute, Chattnag Road, Allahabad 211 019, India}
\affiliation{Homi Bhabha National Institute, Training School Complex, Anushakti Nagar,
     Mumbai 400085, India}
\author{Mehedi Masud}
\email{jomiye.adda@gmail.com}
\affiliation{Institute of Physics, Sachivalaya Marg, Sainik School Post, 
Bhubaneswar 751005, India}
\affiliation{Homi Bhabha National Institute, Training School Complex,
Anushakti Nagar, Mumbai 400085, India}
    \author{Mary Bishai}
  \email{mbishai@bnl.gov}
\affiliation{Brookhaven National Laboratory,  P.O. Box 5000, Upton, NY 11973, USA}
\author{Poonam Mehta}
\email{pm@jnu.ac.in}
\affiliation{School of Physical Sciences, Jawaharlal Nehru University, 
      New Delhi 110067, India}

%\maketitle

%\title{Enhancing the sensitivity to CP violation, mass hierarchy and 
%octant of $\theta_{23}$ with experimentally tunable  beams at DUNE}          
\date{\today}
%===========================================================================
\begin{abstract}
Even though neutrino oscillations have been conclusively established, there are a few unanswered questions pertaining to leptonic Charge Parity violation (CPV), mass hierarchy (MH) and $\theta_{23}$ octant degeneracy. Addressing these questions is of paramount importance at the current and future neutrino experiments including the Deep Underground Neutrino Experiment (DUNE) which has a baseline of 1300 km.  
In the standard mode, DUNE is expected to run with a {\textit{low energy}} (LE) tuned beam which peaks around the first oscillation  maximum ($2-3$ GeV) (and then sharply falls off as we go to higher energies). However, the wide band nature of the beam available at long baseline neutrino facility (LBNF) allows for the flexibility in utilizing beam tunes that are well-suited at higher energies as well. 
In this work, we utilize a beam that provides high statistics at higher energies which is referred to as the {\textit{medium energy}} (ME) beam. This opens up the possibility of exploring not only the usual oscillation channels but also the $\nu_{\mu} \to \nu_{\tau}$ oscillation channel which was otherwise not accessible.    
Our goal is to find an optimal combination of beam tune and runtime (with the total runtime held fixed) distributed in neutrino and antineutrino mode that leads to an improvement in the sensitivities of these parameters at DUNE. In our analysis, we incorporate all the three channels ($\nu_{\mu} \to \nu_{e}, \nu_{\mu} \to \nu_{\mu}, \nu_{\mu} \to \nu_{\tau}$)  and develop an understanding of their relative contributions in sensitivities at the level of $\Delta \chi^2$. 
Finally, we obtain the preferred combination of runtime using both the beam tunes as well as neutrino and antineutrino mode that lead to enhanced sensitivity to the current unknowns in neutrino oscillation physics {\textit{i.e.}}, CPV, MH and $\theta_{23}$ octant.
\end{abstract}
\maketitle
%=================================================================
\section{Introduction}
%===============================================
Neutrino was first postulated by Pauli in 1930 to tackle the issue of non-conservation of energy in the $\beta$ decay spectrum and subsequently discovered experimentally in 1956~\cite{Cowan:1992xc}.
 The idea of neutrino oscillations as $\nu \to \bar\nu$  oscillations was proposed by Pontecorvo~\cite{Pontecorvo:1957cp,Pontecorvo:1957qd} in 1957 which was thought to be the leptonic analogue of the $K_0 \to \bar K_0$ oscillations in the hadronic sector. However, soon after the discovery of the second type of neutrino ($\nu_\mu$), the idea evolved into neutrino flavor oscillations which was proposed by Maki, Nakagawa and Sakata~\cite{mns} as well as by 
  Gribov and Pontecorvo~\cite{Gribov:1968kq}.  Neutrino oscillations among the three active flavours  imply that at least two of the neutrino states are massive which can not be reconciled within the Standard Model (SM) of particle physics. 
The discovery of neutrino oscillations in multiple experiments involving different energies and baselines was awarded the Nobel Prize for Physics in 2015~\cite{nobel2015}. 
 Neutrino sector offers us with a unique opportunity to explore the mysterious world of physics beyond the Standard Model (BSM). 
 
 Neutrino oscillations are governed by six parameters: three mixing angles  ($\ta, \tb, \tc$), two mass squared differences ($\sdm = m_{2}^{2}-m_{1}^{2}, \ldm = m_{3}^{2}-m_{1}^{2}$) and one Dirac CP phase ($\delta$).
Among the various parameters, $\ta$, $\sdm$ (as well as the sign of $\sdm$) and $\tb$ have been measured quite precisely~\cite{Zyla:2020zbs}. Also, we have a fairly good idea about the magnitude of $\ldm$~\cite{Zyla:2020zbs}. The focus has now shifted to address three key  questions such as what is the value of  the CP phase $\delta$ that enters the oscillation formalism,  what is the sign of $\ldm$ (also referred to as the neutrino mass hierarchy) and what is the octant of $\tc$ ? 
If $\delta$ is found to be different from $0$ or $\pi$, it would imply CP violation (CPV) in the leptonic sector. The answer to this question is crucially linked to a more fundamental and elusive puzzle vis-$\grave{\text{a}}$-vis why is there baryon asymmetry in the Universe?
In order to match the observed baryon asymmetry, Sakharov's conditions~\cite{Sakharov:1967dj} have to be satisfied i.e., (a) Existence of Baryon number violating process, (b) C and CP Violation, and (c) processes out of thermal equilibrium.  
 A seminal work carried out by Fukugita was to invoke the idea of leptogenesis to achieve baryogenesis~\cite{Fukugita:1986hr} (see also \cite{Davidson:2008bu} for a review on leptogenesis). 
 Thus, establishing whether CP is violated in the leptonic sector would provide a key missing ingredient towards solving the mystery of matter-antimatter asymmetry in the observed universe\footnote{See \cite{Branco:2011zb}  for a review on leptonic CPV.}.
 The next question pertains to the determination of mass hierarchy \ie, whether the three neutrino mass eigenstates $m_{1}, m_{2}, m_{3}$ are arranged in normal (NH \ie, $\ldm > 0$) or inverted (IH \ie, $\ldm < 0$) hierarchy is of fundamental importance. 
 Apart from shedding light into the plausible set of models for neutrino mass generation\footnote{For \eg, for models based on flavour symmetry, the ones exhibiting softly broken $L_{e}-L_{\mu}-L_{\tau}$ symmetry predict IH~\cite{Petcov:1982ya} and GUT models employing a type I seesaw mechanism prefers NH. See \cite{King:2014nza} for a recent review of neutrino mass models.}, this will also help in determining the nature of neutrino (i.e., Dirac or Majorana) through neutrinoless double beta decay via the effective Majorana mass parameter $m_{\beta\beta}$~\cite{Haxton:1985am, Elliott:2004hr, Aalseth:2004hb}. 
 %\footnote{\comment{For \eg, an inverted ordering coupled with $|m_{\beta\beta}| \gtrsim 15$ meV will indicate the Majorana nature of neutrinos\cite{Elliott:2004hr}.}}.
 Finally, the close to maximal value of the mixing angle $\tc$ could indicate the presence of a new symmetry, the $\mu-\tau$ symmetry in nature~\cite{Lam:2001fb, Harrison:2002et}. 
The determination of the correct octant of $\tc$ \ie, whether $\tc > \pi/4$ (Higher octant or HO) or $\tc < \pi/4$ (Lower octant or LO) or $\tc = \pi/4$ (maximal mixing), plays an crucial role in validating a certain class of models to generate neutrino mass related to the $\mu-\tau$ symmetry\footnote{See \cite{Xing:2015fdg} for a recent review of the $\mu-\tau$ symmetry in neutrino physics.}.  
 
The available data from the currently running long-baseline neutrino oscillation experiments 
Tokai to Kamioka (T2K)~\cite{Abe:2013hdq} and NuMI Off-axis 
$\nu_e$ Appearance (NO${\nu}$A)~\cite{Ayres:2004js} have 
 started uncovering a few of  the open issues mentioned above.
Latest T2K results~\cite{Abe:2019vii} hint towards HO with 
 $\sin^{2}\theta_{23}$ = $0.53^{+0.03}_{-0.04}$ for both NH
and IH. 
For the first time, T2K has been able to rule out a large 
range of values of $\delta$ around $\pi/2$ at $3\sigma$ 
C.L. irrespective of mass hierarchy. It also excludes CP conservation ($\delta = 0$ or $\pi$) at 95\% C.L. 
The most recent measurements by the 
NO$\nu$A Collaboration~\cite{Acero:2019ksn} using both 
$\nu$ and $\bar{\nu}$ running mode hint towards NH at $1.9\sigma$ C.L. and shows a weak preference for 
$\theta_{23}$ lying in HO at a C.L. of $1.6\sigma$.
The NO$\nu$A data excludes most of the choices near 
$\delta = \pi/2$ for IH at a C.L. $\geqslant 3\sigma$.
These results are expected to be further strengthened with as more data becomes available. 
The recent global analyses of the available neutrino data~\cite{deSalas:2020pgw, globalfit, Capozzi:2018ubv, Esteban:2018azc} also indicate preference for NH at more than $3\sigma$ C.L. and  a non-maximal $\theta_{23}$ around $2\sigma$ 
with a slight preference for HO.  
At the same time, caution needs to be exercised in interpretation of results of T2K and NO$\nu$A experiments in resolving mass hierarchy and CPV~\cite{Tortola:2020ncu, Kelly:2020fkv}. It is of importance to address the key questions listed above and  measure the unknowns in an unambiguous manner. 

The upcoming Deep Underground Neutrino Experiment (DUNE)~\cite{Acciarri:2015uup, Abi:2020evt, Abi:2020qib} has the potential to resolve the key questions mentioned above with a very high precision.  DUNE is expected to use the standard low energy (LE) tuned flux (having a peak around $2-3$ GeV and sharply falling at energies $E \gtrsim 4$ GeV) with a total runtime of 7 years distributed equally between the $\nu$ and $\bar{\nu}$ modes (3.5 years + 3.5 years). 
Among the viable additional beams that can be used at DUNE, 
%there exist a medium energy (ME) tuned flux which offers substantial statistics even at energies $E \gtrsim$ 4 GeV (albeit at the cost of some loss of statistics around $2-3$ GeV). 
%
there is a possibility of deploying a medium energy tune (ME) based on the NuMI focusing system which offers substantial statistics even at energies $E \gtrsim$ 4 GeV (albeit at the cost of some loss of statistics around $2-3$ GeV). The role of the ME beam at DUNE as been explored in disentangling non-standard neutrino interactions (NSI) from the standard oscillation~\cite{Masud:2017bcf}, constraining parameter degeneracies in the presence of NSI~\cite{Masud:2018pig} and constraining unitarity using the $\nu_\mu \to \nu_\tau$ channel~\cite{deGouvea:2019ozk,Ghoshal:2019pab}.

To exploit the full potential of DUNE, we make use of this ME beam with a focus to exploit the high statistics it offers at higher energies. Since the neutrinos and antineutrinos encounter different potential due to earth matter effects, the variation of runtime of a long baseline experiment such as DUNE while running in neutrino ($\nu$) mode versus antineutrino ($\bar\nu$) mode could lead to a difference in sensitivities to MH, CPV and octant of $\theta_{23}$.
In the present work, we combine LE and ME beam tunes and vary  runtime in the $\nu$ and $\bar{\nu}$ modes corresponding to each of the beams with the goal to improve the sensitivities of DUNE to MH, CPV and octant of $\theta_{23}$. 

Let us summarize some previous studies which focus on the variation of runtime between the $\nu$ and $\bar{\nu}$ modes.  In \cite{Huber:2009cw, Agarwalla:2013ju, Machado:2013kya, Ghosh:2014zea, Ghosh:2015ena}, the authors  have carried out optimization of runtime combinations in $\nu$ and $\bar{\nu}$ mode for the currently running long baseline experiments, NO$\nu$A and T2K. They demonstrate that the degeneracies are better resolved with particular choice of runtime combinations. 
In the context of DUNE, the authors of~\cite{Ghosh:2014rna, Nath:2015kjg} demonstrate that improved sensitivities to CPV, MH and octant of $\tc$ can potentially be reached for particular combinations of runtime in $\nu$ and $\bar{\nu}$ mode. 
\cite{Ghosh:2014rna} further discusses the possibility of a combined analysis with T2K and NO$\nu$A, while \cite{Nath:2015kjg} illustrates the importance of $\bar{\nu}$ runtimes at DUNE. 
\cite{Coloma:2014kca} discusses about different runtime combinations that can give better precision in measuring the CP violating phase and the octant  of $\tc$ at the erstwhile LBNE~\cite{Adams:2013qkq} (which was  the predecessor of DUNE). 
\cite{Ballett:2016daj} gives a detailed account of various possible optimized configurations of DUNE, including an analysis of  total cumulative runtime.  A study has been carried out with the variation of runtime using only the LE flux in the $\nu$ and $\bar{\nu}$ modes~\cite{bnl_le}.

The present work goes beyond studies existing in the literature in two aspects. 
 Firstly, we use an additional beam tune  (ME beam) in conjunction with the standard LE beam (thus utilising the wide band nature of DUNE to a greater extent) and  analyze the variation of runtime for both the fluxes in the $\nu$ and $\bar{\nu}$ modes in order to improve the sensitivities. 
Secondly, since the ME beam we implement in our simulation has been optimized to detect a large number of $\nu_{\tau}$ events, we include a $\nutau$ (as well as $\nutaubar$) appearance channel, in addition to $\nue$ ($\nuebar$) and $\numu$ ($\numubar$) channels, in our analysis to estimate the sensitivities to CPV, MH and octant of $\tc$. 

This article is organised as follows. 
 In Sec.\ \ref{sec:basics}, we give briefly describe effective Hamiltonian and the parameters governing standard neutrino oscillations. Sec.\ \ref{sec:beam} discusses beam tunes i.e., the LE and ME fluxes we have used in our analysis.  Our analysis methodology is discussed in Sec.\ \ref{sec:analysis}.  In Sec.\ \ref{sec:opt}, we show our main results, the optimized runtime combinations that generate the best sensitivities to CPV, MH and  octant of $\tc$.
 Finally, we summarize our results and conclude. 
 In the Appendix, we discuss the role of the three oscillation channels to resolve the questions of CPV, MH and octant of $\tc$ in the level of probabilities.
 %============================
\section{Effective Hamiltonian in the Standard oscillation framework}
\label{sec:basics}
%============================
The effective Hamiltonian describing neutrino propagation in the flavour basis is expressed as, 
 \bea
 \label{hexpand} 
 {\mathcal
H}^{}_{\mathrm{f}} &=&   {\mathcal
H}^{}_{\mathrm{v} } +  {\mathcal
H}^{}_{\mathrm{SI} } 
\nonumber 
\\
&
=&\frac{\ldm}{2E} \left\{ {\mathcal U} \left(
\begin{array}{ccc}
0   &  &  \\  &  \alpha &   \\ 
 &  & 1 \\
\end{array} 
\right) {\mathcal U}^\dagger  + A   \left(
\begin{array}{ccc}
1  & 0 & 0 \\
0 &  0 & 0  \\ 
0 & 0 & 0 \\ 
\end{array} 
\right)  \right\}  \ ,
 \eea 
where $\alpha = {\sdm}/{\ldm}$ and $A = {2 \sqrt{2} E G_F n_e}/{\ldm}$. 
$2 \sqrt{2} E G_F n_e$ is the standard charged current (CC) potential due to
the coherent forward scattering of neutrinos propagating through a medium of electron density $n_{e}$, $G_{F}$ being the Fermi constant.

${\mathcal
  U}$ is the three flavour neutrino mixing matrix and is responsible for 
diagonalizing the vacuum part (${\mathcal H}_{\mathrm{v}}$) of the Hamiltonian. 
It is parameterized by three angles $\theta_{12},\theta_{23},\theta_{13}$ and 
one  phase  $\delta$~\footnote{In the general case of $n$ flavors the leptonic mixing matrix 
${\mathcal U}$ depends on $(n-1)(n-2)/2$ Dirac-type 
CP-violating  phases. If the neutrinos are Majorana particles, there are $(n-1)$ additional, 
so called Majorana-type CP-violating phases.}. If neutrinos are Majorana particles, there can be 
two additional Majorana-type phases in the three flavour case but they are of no consequence 
 in neutrino oscillations.  
 In the commonly used Pontecorvo-Maki-Nakagawa-Sakata (PMNS)
 parametrization~\cite{Zyla:2020zbs}, ${\cal U}$ is given by
\begin{eqnarray}
{\mathcal U}^{} &=& \left(
\begin{array}{ccc}
1   & 0 & 0 \\  0 & c_{23}  & s_{23}   \\ 
 0 & -s_{23} & c_{23} \\
\end{array} 
\right)   
  \left(
\begin{array}{ccc}
c_{13}  &  0 &  s_{13} e^{- i \delta}\\ 0 & 1   &  0 \\ 
-s_{13} e^{i \delta} & 0 & c_{13} \\
\end{array} 
\right)  \left(
\begin{array}{ccc}
c_{12}  & s_{12} & 0 \\ 
-s_{12} & c_{12} &  0 \\ 0 &  0 & 1  \\ 
\end{array} 
\right)  \ ,
\label{u}
 \end{eqnarray}
where $s_{ij}=\sin {\theta_{ij}}, c_{ij}=\cos \theta_{ij}$.

%===============================
\section{Neutrino Beam Tunes and Event spectra at DUNE}  
\label{sec:beam}
%==============================
\begin{figure}[htb!]
\centering
\includegraphics[width=.5\textwidth]
{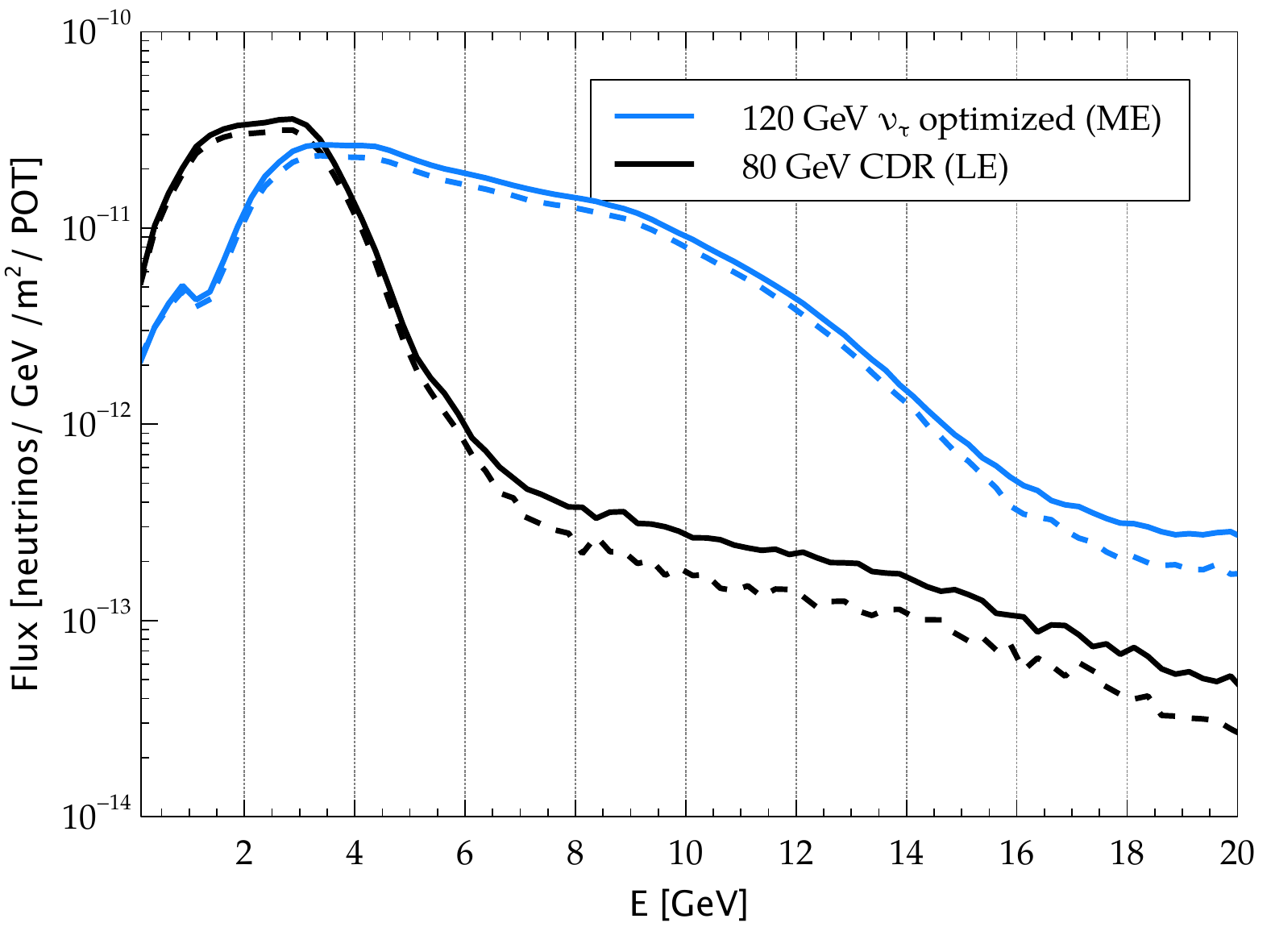}
\caption{\footnotesize{Comparison of the different beam tunes - the 80 GeV flux used in DUNE Conceptual Design Report (CDR) is referred to as LE flux while 120 GeV, $\nu_\tau$ optimized flux is referred to as the ME flux. The solid (dashed) curves represent the $\nu_{\mu}$ ($\bar{\nu}_{\mu}$) flux. 
}}
\label{fig:flux}
\end{figure}
%%%%%%%%%%%%%%%% 
 The simulations have been carried out using the widely used General Long Baseline Experiment Simulator (GLoBES)~\cite{Huber:2004ka,Huber:2007ji} which solves the full three flavour neutrino propagation equations numerically using the Preliminary Reference Earth Model (PREM)~\cite{Dziewonski:1981xy} 
%%%%%
%%%
density profile of the Earth\footnote{We use the matter density as given by PREM model.
 In principle, we can allow for uncertainty in the Earth matter density in our calculations but  it would not impact our results drastically~\cite{Gandhi:2004md, Gandhi:2004bj, Kelly:2018kmb, Chatterjee:2018dyd}.}. The most recent DUNE configuration files from the DUNE collaboration~\cite{Alion:2016uaj} have been used.
 A total runtime  of 7 years with an on-axis 40 kiloton liquid argon far detector 
(FD) housed at the Homestake Mine in South Dakota over a baseline of 1300 km has been incorporated in the simulations.

 We use two broad-band beam tunes : (i) the standard LE beam tune used in DUNE CDR~\cite{Alion:2016uaj} and (ii) the ME beam tune optimized for $\nu_{\tau}$ appearance.~%\cite{x}. 
  The beams are obtained from a G4LBNF 
simulation~\cite{Agostinelli:2002hh,Allison:2006ve} of the LBNF 
beam line using NuMI-style focusing. 
These two broad-band beam tunes  are consistent with what could be achieved by the LBNF facility and we illustrate their comparison in Fig.~\ref{fig:flux}. {\hlpm{
We have chosen a design for the ME beam that is nominally compatible with the space and infrastructure capabilities of the LBNF/DUNE beamline and that is based on a real working beamline design currently deployed in NuMI/NOvA.}}
 The beamline parameters assumed for the different design fluxes  used 
in our analyses are listed in Table~\ref{tab:cdr}. 
%%%%%%%%%%%%%%%%%%%%%%%
\begin{table}[htb!]
\centering
{{
\begin{tabular}{ l  l  l}
\hline
&&\\
Parameter & LE (CPV optimized design) & ME ($\nu_{\tau}$ optimized) \\
&&\\
\hline
&&\\
Proton Beam energy  & 80 GeV & 120 GeV \\
Proton Beam power & 1.07 MW & 1.2 MW\\ 
Protons on target (POT) per year & $1.47 \times 10^{21}$ & $1.10 \times 10^{21}$\\
Focusing &  3 horns & 2 NuMI horns \\
 & genetic optimisation & 17   m apart\\
%Target location & embedded in horn 1 & -2 m \\ 
Horn Current & 294 kA & 230 kA \\
Decay pipe length & 194   m & 200 m \\
Decay pipe diameter & 4 m & 4 m  \\
&&\\
\hline
\end{tabular}
}}
\caption{\label{tab:cdr}  
Beamline parameters assumed for the different design fluxes used 
in our sensitivity calculations~\cite{Acciarri:2015uup, Alion:2016uaj, Abi:2020evt}. The 
 target is a thin graphite cylinder 2 interaction lengths long.} 
\end{table}
%%%%%%%%%%%%%%%%%%%%%%%%%

%%%%%%%%%%%%%%%%%%%%%%
\begin{figure}[ht!]
\centering
\includegraphics[width=.7\textwidth]
{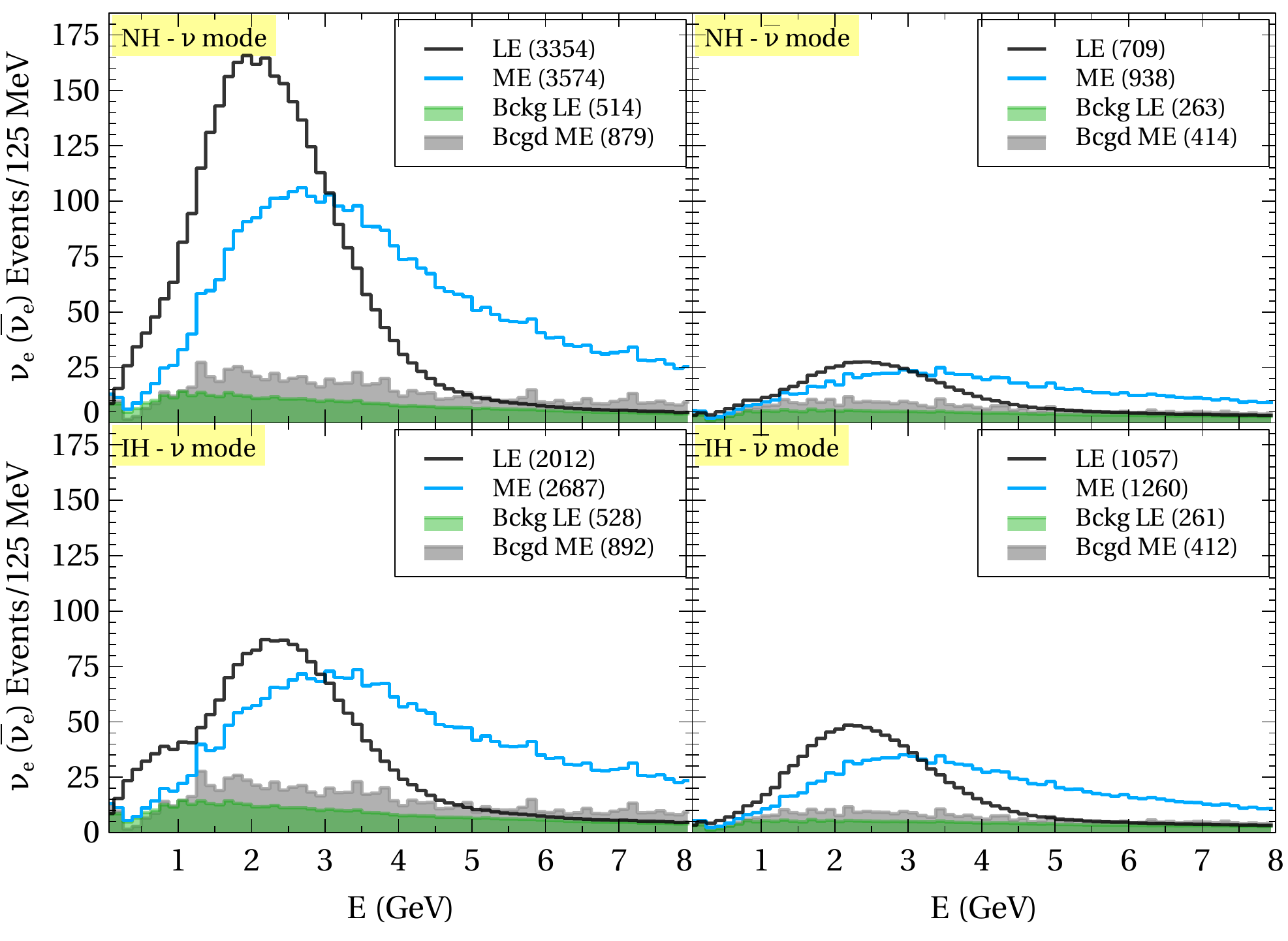}
\caption{\footnotesize{\hlpm{The $\nue$ event spectra at DUNE are shown for different cases - the top (bottom) row corresponds to NH (IH) while the left (right) column corresponds to $\nu$ ($\bar \nu$) mode.
The events  obtained using the  standard LE beam (the $\nu_{\tau}$-optimized ME beam)  are depicted using black (blue) curves in each of the four panels.
The total background corresponding with LE (ME) beam  are  shown as green (grey) shaded regions. 
In generating each of the event/background spectra, a total runtime of 7 years have been used.  The numbers in the parentheses in the legends are the corresponding total number of events summed over the energy bins upto 8 GeV. The best-fit values of the oscillation parameters  have been used to generate these event spectra (see  Table~\ref{tab:parameters}).}}}
\label{fig:event1}
\end{figure}

\begin{figure}[htb!]
\centering
\includegraphics[width=.7\textwidth]
{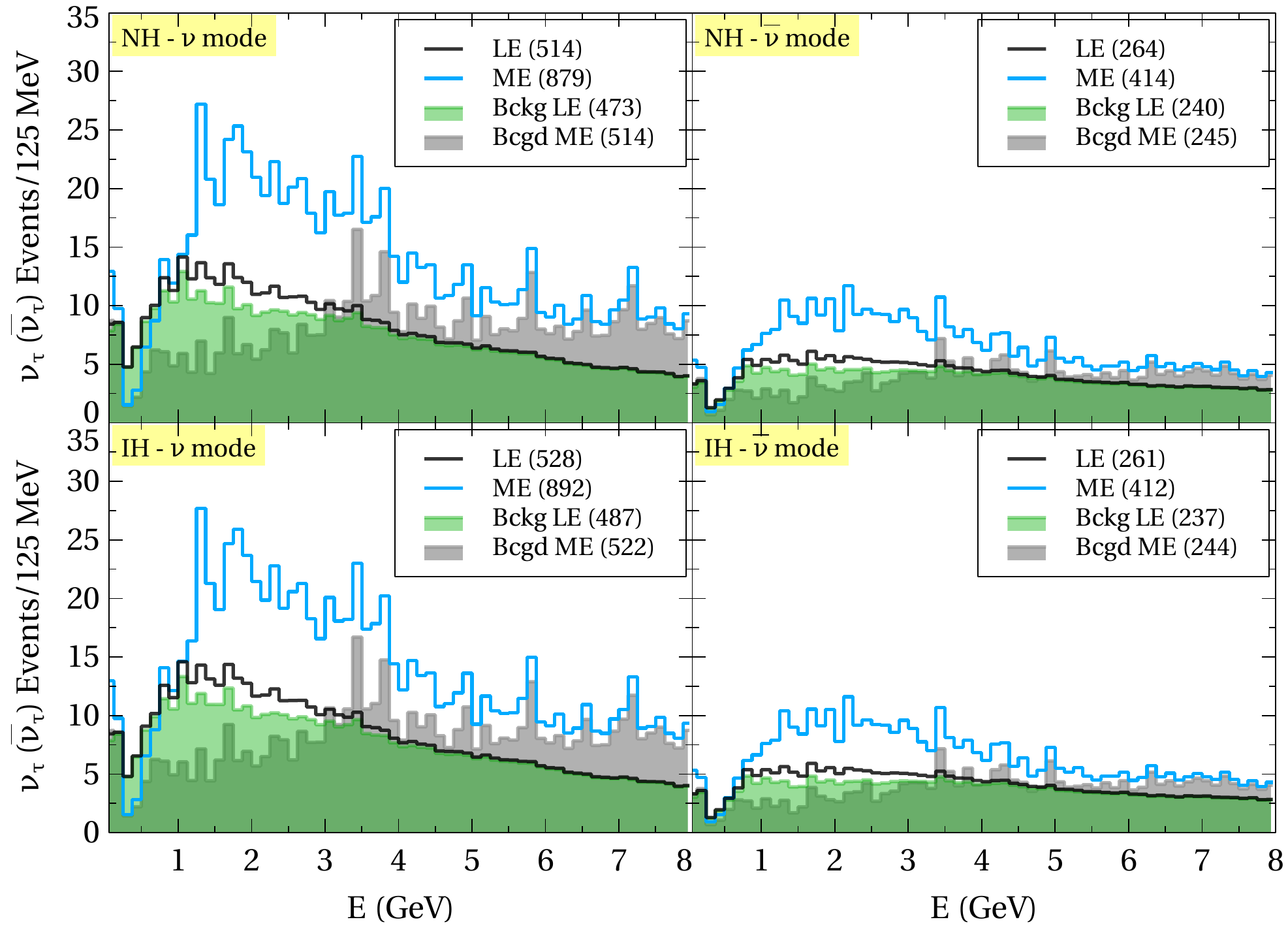}
\caption{\footnotesize{\hlpm{The $\nutau$ event spectra at DUNE are shown for different cases - the top (bottom) row corresponds to NH (IH) while the left (right) column corresponds to $\nu$ ($\bar \nu$) mode.
The events  obtained using the  standard LE beam (the $\nu_{\tau}$-optimized ME beam)  are depicted using black (blue) curves in each of the four panels.
The total background corresponding with LE (ME) beam  are  shown as green (grey) shaded regions. 
In generating each of the event/background spectra, a total runtime of 7 years have been used.  The numbers in the parentheses in the legends are the corresponding total number of events summed over the energy bins upto 8 GeV. The best-fit values of the oscillation parameters  have been used to generate these event spectra (see  Table~\ref{tab:parameters}).}}}
\label{fig:event2}
\end{figure}

{\hlpm{In Fig.\ \ref{fig:event1}, 
we show the $\nue$ (and $\nuebar$) event spectra at DUNE when the beam tune is either the standard LE beam (black) or the $\nu_{\tau}$-optimized ME beam (blue).  }}
%The solid (dotted) histogram refers to the running of DUNE in $\nu$ ($\bar{\nu}$) mode.
In order to make a fair comparison, the total runtime has been held fixed to 7 years  in generating each of these four spectra. 
% The left (right) panel corresponds to the case of NH (IH). 
We note that though the LE flux gives an excess of events around $E \lesssim 3$ GeV, the ME beam offers complementarity by generating more events beyond $E \gtrsim 3$ GeV. 
In fact, for the chosen values of the parameters, the total number of events summed over the energy bins upto 8 GeV (as indicated in the figure), is slightly more for the ME beam with respect to the LE beam\footnote{If the parameters are allowed to vary within the $3\sigma$ allowed range, it is found that this conclusion holds in much of the parameter space.}. 
This trend is visible for both $\nu$ and $\bar{\nu}$ modes irrespective of the choice of the hierarchy. 
In view of the above stated observations,  we investigate whether a combination of LE and ME beam could lead to  an improvement in the sensitivities to CPV, MH and octant of $\tc$  over and above what is expected from the LE beam alone.   
{\hlpm{In  Fig.~\ref{fig:event2}, we show the $\nutau$ (and $\nutaubar$)  event spectra at DUNE when the beam tune is either the standard LE beam (black) or the $\nu_{\tau}$-optimized ME beam (blue).  It is worth mentioning that ME beam opens up the possibility to exploit this channel in addition to the $\nue$ channel which further aids in improving the outcome of the sensitivity studies.}}

{\hlpm{For classification of events using the two beam tunes, we follow Alion et. al.~\cite{Alion:2016uaj}. In the GLoBES configurations~\cite{Alion:2016uaj}, event classification was listed in distinct categories including electron neutrino appearance signal, $\nue$ (CC), muon neutrino charged current signal, $\numu$ (CC)  as well as  neutrino neutral current (NC) background, $\nu_\mu / \nu_e \to X$ (NC) both in appearance and disappearance modes. It should be noted that tau neutrino appearance  $\nutau$ (CC) was included in~\cite{Alion:2016uaj},  but as a background. The corresponding systematics/efficiencies were provided as supplementary files in~\cite{Alion:2016uaj}.
In the present study, we have incorporated a new category for tau neutrino appearance signal, $\nutau$ (CC). 
The backgrounds corresponding to this category include $\nue$ (CC), $\numu$ (CC) (due to the leptonic decay of tau lepton with branching fraction $\sim$ 35\%) as well as NC background (due to the hadronic decay of the tau lepton with branching fraction  $\sim$ 65\%). Contamination due to wrong sign leptons is also taken into account as  background.  }}

%%%%%%%%%%%%%%%%%%%%%%
%\begin{figure}[htb!]
%\centering
%\includegraphics[width=.7\textwidth]
%{event.pdf}
%\caption{\footnotesize{The $\nue$ event spectra at DUNE are shown for the standard LE beam (black) only and for the $\nu_{\tau}$-optimized ME beam (blue). The solid (dotted) lines correspond to running in $\nu$ ($\bar{\nu}$) mode. 
%The total backgrounds corresponding to LE-$\nu$ mode and ME-$\nu$ mode are also shown with the green and grey shaded area respectively. 
%In generating each of the event/background spectra, a total runtime of 7 years have been used. The left (right) column depicts the case of NH (IH). The numbers in the parentheses in the legends are the corresponding total number of events summed over the energy bins upto 8 GeV. The best-fit values of the oscillation parameters  have been used to generate these event spectra (see  Table~\ref{tab:parameters}).}}
%\label{fig:event}
%\end{figure}

%%%%%%%%%%%%%%%%%%%%%%

%=============================
\section{Analysis method}
 \label{sec:analysis}
 %==============================
 \begin{table}[h]
\centering
\scalebox{0.9}{
\begin{tabular}{| c | c | c | c |}
\hline
&&&\\
Parameter & Best-fit-value & 3$\sigma$ interval & $1\sigma$ uncertainty  \\
&&&\\
\hline
&&&\\
%{\sl{SI}} &&\\
%&&\\
$\theta_{12}$ [Deg.]             & 34.3                    &  31.4 - 37.4   &  2.9\% \\
$\theta_{13}$ (NH) [Deg.]    & 8.58              &  8.16  -  8.94   &  1.5\% \\
$\theta_{13}$ (IH) [Deg.]    & 8.63              &  8.21  -  8.99   &  1.5\% \\
$\theta_{23}$ (NH) [Deg.]        & 48.8                     &  41.63  - 51.32    &  3.5\% \\
$\theta_{23}$ (IH) [Deg.]        & 48.8                     &  41.88  - 51.30    &  3.5\% \\
$\sdm$ [$\text{eV}^2$]  & $7.5 \times 10^{-5}$  &  [6.94 - 8.14]$\times 10^{-5}$  &  2.7\% \\
$\ldm$ (NH) [$\text{eV}^2$] & $+2.56 \times 10^{-3}$   &  [2.46 - 2.65] $\times 10^{-3}$ &  1.2\% \\
$\ldm$ (IH) [$\text{eV}^2$] & $-2.46 \times 10^{-3}$  & -[2.37 - 2.55]$\times 10^{-3}$  &  1.2\% \\
$\delta$ (NH) [Rad.]   & $-0.8\pi$   & $[-\pi, 0]  \cup [0.8\pi, \pi]$ &  $-$ \\
$\delta$ (IH) [Rad.]   & $-0.46\pi$   & $[-0.86\pi, -0.1\pi]$   & $-$  \\
&&&\\
\hline
\end{tabular}}
\caption{\label{tab:parameters}
 Standard oscillation parameters and their uncertainties used in our study. The values were taken from the global fit analysis in \cite{deSalas:2020pgw}. 
 If the $3\sigma$ upper and lower limit of a parameter is $x_{u}$ and $x_{l}$ respectively, the $1\sigma$ uncertainty is $(x_{u}-x_{l})/3(x_{u}+x_{l})\%$~\cite{Abi:2020evt}.}
\end{table}
 To estimate the sensitivities of DUNE to mass hierarchy (MH), CP violation (CPV) and octant of $\tc$, we perform the standard $\chisq$ analysis. 
 Even though all results are produced numerically with the help of GLoBES software, in order to gain insight, let us  examine the analytical form of the $\chisq$\,\footnote{This is the {\it{Poissonian}} definition of $\chisq$, which in the limit of large sample size, reduces to the Gaussian form.}.
 \begin{align}
\label{eq:chisq}
\Delta \chi^{2}(p^{\text{true}}) = \underset{p^{\text{test}}, \eta}{\text{Min}} \Bigg[&2\sum_{y}^{\text{flux}}\sum_{x}^{\text{mode}}\sum_{j}^{\text{channel}}\sum_{i}^{\text{bin}}\Bigg\{
N_{ijxy}^{\text{test}}(p^{\text{test}}; \eta) - N_{ijxy}^{\text{true}}(p^{\text{true}})
+ N_{ijxy}^{\text{true}}(p^{\text{true}}) \ln\frac{N_{ijxy}^{\text{true}}(p^{\text{true}})}{N_{ijxy}^{\text{test}}(p^{\text{test}}; \eta)} \Bigg\}  \nonumber \\
&+ \sum_{l}\frac{(p^{\text{true}}_{l}-p^{\text{test}}_{l})^{2}}{\sigma_{p_{l}}^{2}}
+ \sum_{k}\frac{\eta_{k}^{2}}{\sigma_{k}^{2}}\Bigg],
\end{align}
where $N^{\text{true}}$ and $N^{\text{test}}$ are the set of true and test events respectively. 
Index $i$ is summed over the energy bins in the range $0-20$ GeV\footnote{We have a total of 71 energy bins in the range 0-20 GeV: $64$ bins each having a width of 
$0.125$ GeV in the energy range of 0 to 8 GeV and $7$ bins
with variable widths beyond $8$ GeV~\cite{Alion:2016uaj}.}. 
The indices $j$ and $x$ are summed over the channels ($\nue, \numu, \nutau$) and the modes ($\nu$ and $\bar{\nu}$) respectively. 
The sum over the index $y$ takes into account the multiple fluxes (LE and ME tuned beams), whenever multiple fluxes are used. 
The term inside the curly braces in Eq.\ \ref{eq:chisq} is the statistical part of  $\chisq$. 
The term $(N^{\text{test}} - N^{\text{true}})$ takes into account the algebraic difference while the third term inside the curly braces considers the fractional difference  between the {\it{test}} and {\it{true}} sets of events. 
$p^{\text{true}}$ and $p^{\text{test}}$ are the set of true and test oscillation parameters respectively and $\sigma_{p_{l}}$ is the uncertainty in the prior measurement of the parameter $p_{l}$. 
The values of the true or best-fit oscillation parameters and their uncertainties as used in the present analysis are tabulated in Table~\ref{tab:parameters}. 
The index $l$ is summed over the number of test oscillation parameters to be marginalized. 
This is known as the {\it{prior}} term. 
The index $k$ is summed over the number of systematics/nuisance parameters present.
This way of treating the nuisance parameters in the $\chisq$ calculation is known as the {\it{method of pulls}}~\cite{Huber:2002mx,Fogli:2002pt,GonzalezGarcia:2004wg,Gandhi:2007td}. 
Regarding the systematics, the $\nu_{e}$ and $\bar{\nu}_{e}$ signal modes have a normalization uncertainties of $2\%$ each, whereas the $\nu_{\mu}$ and $\bar{\nu}_{\mu}$ signals have a normalization uncertainty of $5\%$ each. The $\nu_{\tau}$ and $\bar{\nu}_{\tau}$ signals have a normalization uncertainties of $20\%$ each. 
The background normalization uncertainties vary from $5\%-20\%$ and include correlations among various sources of background (coming from beam $\nu_{e}/\bar{\nu}_{e}$ contamination, flavour misidentification, NC and $\nu_{\tau}$).  
The final estimate of $\chisq$ which is thus a function of the true values of the oscillation parameters, is obtained after a minimization (\ie, marginalization over the $3\sigma$ range of values) over the set of test parameters ($p^{\text{test}}$) and the set of systematics ($\eta$). 
Technically this $\chisq$ is the frequentist method 
of hypotheses testing~\cite{Fogli:2002pt, Qian:2012zn}.

For calculating the sensitivity to MH, we marginalise the test $\ldm$ in the opposite hierarchy in the $3\sigma$ range of values. 
Test parameters $\tc, \tb$ are marginalized over their $3\sigma$ ranges, while the CP phase $\delta$ is marginalized over the full range of $[-\pi, \pi]$. 
For the sensitivity to CPV, the test $\delta$ is allowed to marginalize over only the CP conserving values of 0 and $\pi$ while the true $\delta$ can take any value in the range $[-\pi, \pi]$. 
The other test parameters in that case ($\ldm, \tb, \tc$) are marginalized over the $3\sigma$ range. 
For the calculation of the sensitivity to $\tc$ octant, the test $\tc$ is marginalized in the opposite octant. 
The marginalization of the test parameters $\ldm$ and $\tb$ are carried out over their respective $3\sigma$ range, while that of test $\delta$ is done in the whole allowed range of $[-\pi, \pi]$. 
 
 We keep the total runtime at DUNE fixed at 7 years and numerically calculate $\chisq$ as a function of true $\delta \in [-\pi, \pi]$ after varying the distribution of runtime (with a stepsize of 0.5 year) among the following four variables  :
 
\begin{itemize}
\item
Runtime using LE beam and in neutrino mode ($\len$)

\item
Runtime using LE beam and in anti-neutrino mode ($\lenbar$)

\item
Runtime using ME beam and in neutrino mode ($\men$)

\item
Runtime using ME beam and in anti-neutrino mode ($\menbar$).
\end{itemize}

Note that, since $\len + \lenbar + \men + \menbar = 7$ (years), only three of the above variables are independent. In order to figure out the optimized runtime combination,  we define that combination of $\len, \lenbar, \men, \menbar$ which gives the largest area under the sensitivity curve in the ($\chisq$-true $\delta$) plane for all three unknowns : CPV, MH and octant of $\tc$.
We  refer to the optimized runtime combination estimated in this manner as $\area{CPV}, \area{MH}$ or $\area{OCT}$ respectively for the three quantities.
 Additionally, for the CPV case, we define another optimized combination of runtimes (\ie, $\len, \lenbar, \men, \menbar$) that resolves CPV above $3\sigma$ for  the largest fraction of true $\delta$ parameter space. We refer to this optimized combination as $\fraction{CPV}$.
%=======================================
\section{Optimized runtime combinations : Sensitivity to CPV, MH and octant of $\tc$}
\label{sec:opt}
%=======================================
We present our main results by estimating the optimized runtime combinations of $(\len, \lenbar, \men, \menbar)$  that give the best sensitivities to resolve CPV, MH  and octant of $\tc$. 
In what follows, we obtain the optimal combinations of runtimes and which are reported as $(\len + \lenbar + \men + \menbar)$ for the three different questions addressed in the present work. 
%%%%%%%%%%%%%%%%%%%%%
\begin{figure}[htb]
\centering
\includegraphics[scale=0.5]{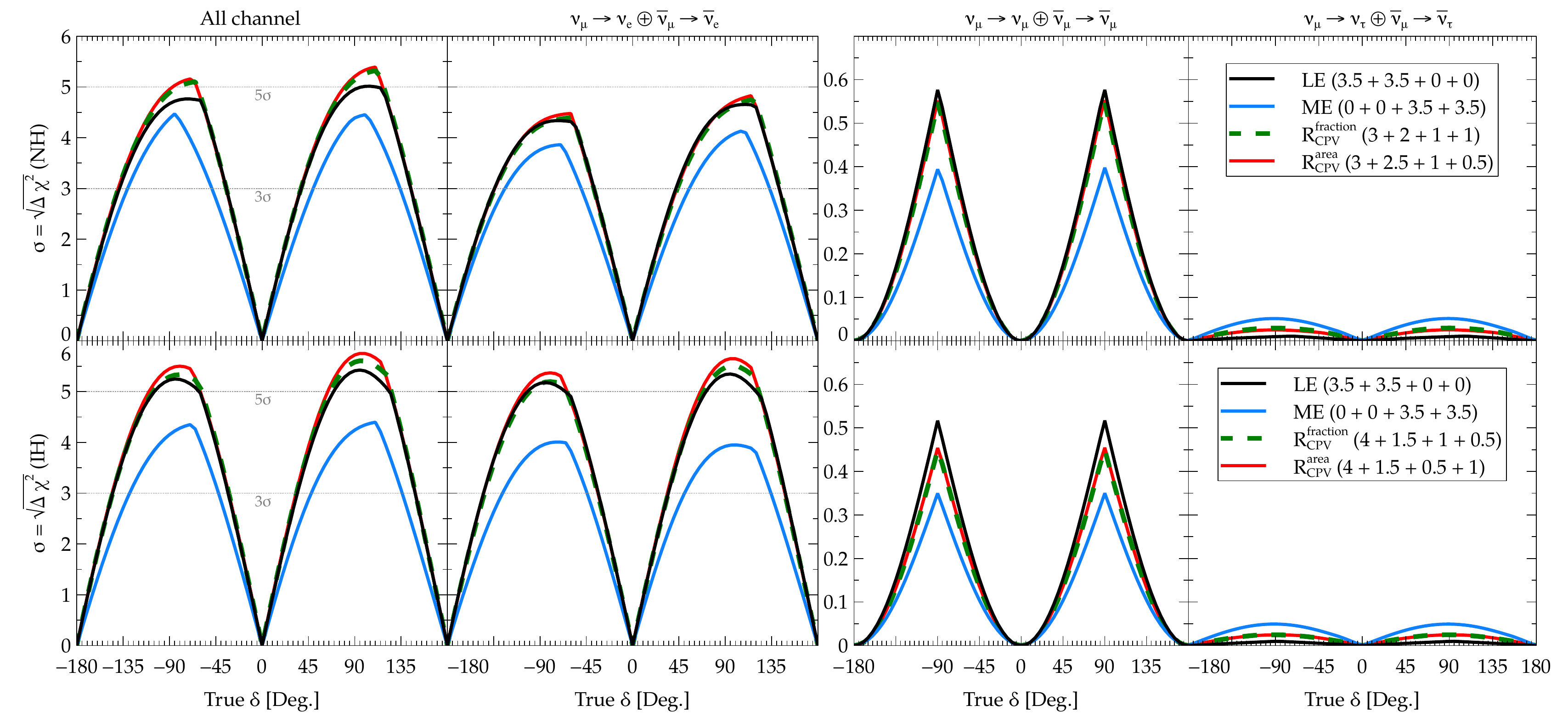}
\caption{\footnotesize{The sensitivity to CPV at DUNE are shown with the variation of true $\delta$, for optimized runtime combination for LE+ME, - found either by maximizing the area under the sensitivity curve (solid red) or by maximizing the fraction of true $\delta$ space for which the sensitivity is above $3\sigma$ (dashed green). 
The sensitivities  obtained for LE only(black) or ME only (blue) beam with equal runtime in $\nu$ and $\bar{\nu}$ modes are also shown. 
The legends signify the runtime combination for each case in the form ($\len + \lenbar + \men + \menbar$). 
The top (bottom) row depicts the case of true NH (IH).
The first column shows the combined results considering all channels while the second, third and fourth column shows the contributions from the individual channels $\nue, \numu$ and $\nutau$ respectively. 
}}
\label{fig:opt_cp}
\end{figure}
%%%%%%%%%%%%%%%%%%%%%%%%

{\underline{\bf{Sensitivity to CP violation :-}}} In Fig.\ \ref{fig:opt_cp}, we illustrate the optimized runtime combination of $(\len + \lenbar + \men + \menbar)$ that give the best  sensitivity to CPV at DUNE by 
  (a) maximizing the area under the sensitivity curve (red solid), and 
  (b) maximizing the fraction of true $\delta$ parameter space that resolves the CPV sensitivity above $3\sigma$ (green dashed).
  For comparison, we also show the CPV sensitivities by using only the standard LE beam (black) or the ME beam only (blue) -  the distribution of runtime 
  being equally (3.5 years) shared between the $\nu$ and $\bar{\nu}$ modes for each case.
  The first column shows the main results where (neutrino and antineutrino) contributions from all three oscillation channels ($\nue, \numu, \nutau$) are considered. 
  The second, third and fourth column shows the (neutrino and antineutrino) contribution from the individual channels  $\nue, \numu$ and $\nutau$ respectively. 
  The top (bottom) row depicts the case of true NH (IH). 

  For NH, the sensitivity to CPV at the best runtime combination ($\area{CPV}$ or $\fraction{CPV}$) is enhanced beyond $5\sigma$ around the CP violating values ($\approx \pm \pi/2$) of true $\delta$.  
  The $5\sigma$ reach of the sensitivity to CPV was otherwise not achievable using only the standard LE beam or the ME beam with a runtime of (3.5 years $\nu$ mode + 3.5 years  $\bar{\nu}$ mode for either of the beams). 
 For the NH scenario, an optimized runtime combination $(\area{CPV} \equiv {\bf{3+2.5+1+0.5)}}$  estimated by maximizing the {\it{area}} of  implies that DUNE needs to run for a total of 
  (a) 5.5 years using the LE beam (with 3 years in $\nu$ mode and 2.5 years\ in $\bar{\nu}$ mode), and 
  (b) 1.5 years using the ME beam (with 1 year in $\nu$ mode and 0.5 year \ in $\bar{\nu}$ mode). It should be noted that the optimized runtime combination obtained by maximizing the {\it{fraction}} is similar to that obtained while maximizing the area. 
  
  For IH, it is found that  the CP violation sensitivity does not  improve much when we use a combination of beam tunes (LE+ME). This is because  
  the sensitivity generated by LE beam is somewhat higher for IH than  for NH which gives less scope of improvement. 
  
  It is worth mentioning that most of the contribution to the CP violation sensitivity arises from the $\nue$ appearance channel.  In Appendix~{\bf A}, we try to explain why this is the dominant channel while the other two channels (i.e., $\numu$ and $\nutau$) give rise to small contribution (see also \cite{Masud:2015xva,Masud:2016bvp}).
%%%%%%%%%%%%%%%%%%%%%%%
\begin{figure}
\centering
\includegraphics[scale=0.5]{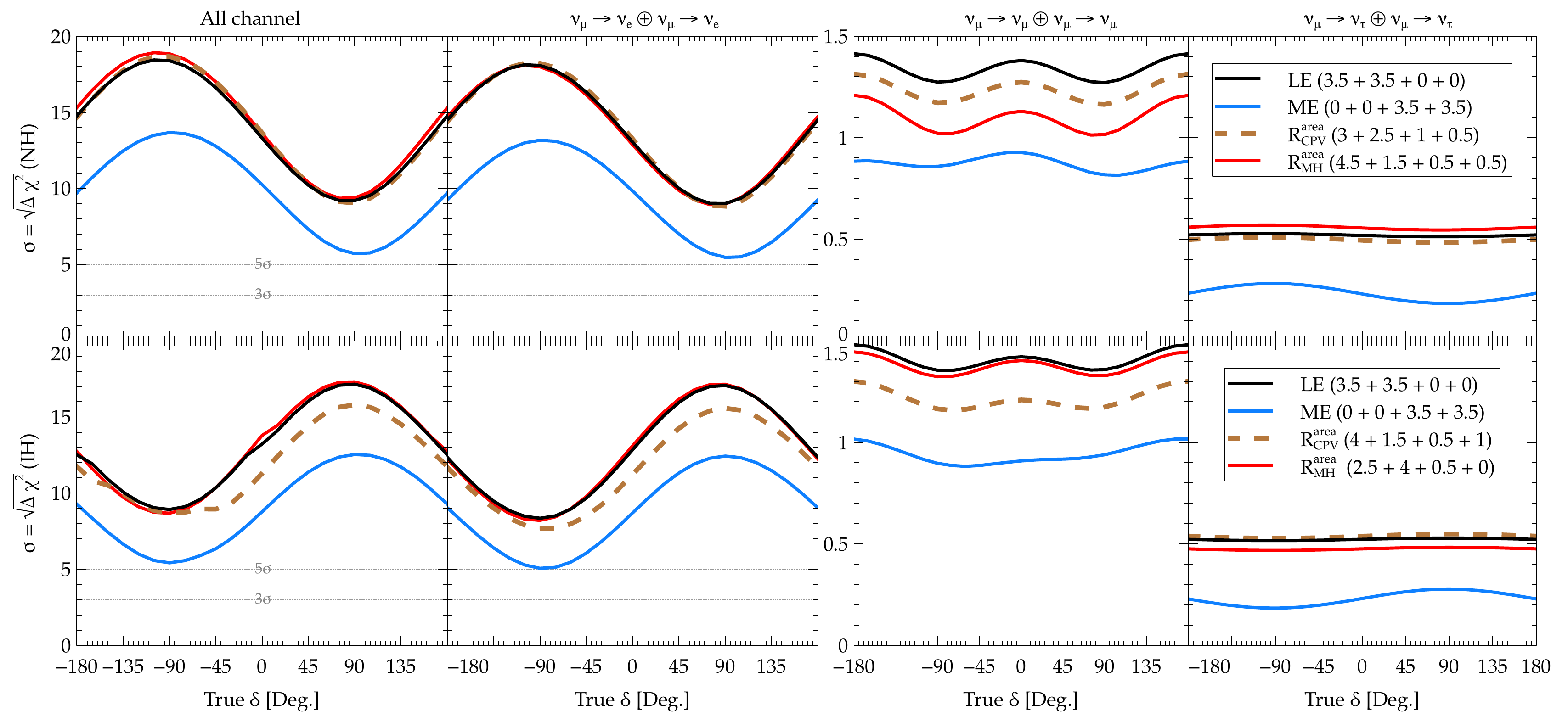}
\caption{\footnotesize{The sensitivities to MH at DUNE are shown with the variation of true $\delta$, for optimized runtime combination for LE+ME, - found by maximizing the area under the sensitivity curve (solid red). 
The sensitivities  obtained for LE only(black) or ME only (blue) beam are also shown with equal runtime in $\nu$ and $\bar{\nu}$ modes. 
The legends signify the runtime combination for each case in the form ($\len + \lenbar + \men + \menbar$). 
For comparison, the MH sensitivities corresponding to $\area{CPV}$ are also shown (dashed brown). 
The legends signify the runtime combination for each case in the form ($\len + \lenbar + \men + \menbar$). 
The top (bottom) row depicts the case of true NH (IH).
The first column shows the combined results considering all channels while the second, third and fourth column shows the contributions from the individual channels $\nue, \numu$ and $\nutau$ respectively.}}
\label{fig:opt_mh}
\end{figure}
%%%%%%%%%%%%%%%%%

{\underline{\bf{Sensitivity to MH :-}}} In Fig.\ \ref{fig:opt_mh}, we illustrate the optimized runtime combination (red curves) of $(\len + \lenbar + \men + \menbar)$ that gives the best MH sensitivity at DUNE. 
For the sake of comparison, we additionally show the MH sensitivities (dashed brown) corresponding to the best runtime combination ($\area{CPV}$) estimated for CPV sensitivity.
We also show the sensitivities to MH by using only the standard LE beam (black) or the ME beam (blue),  the distribution of runtime being equally (3.5 year) shared between the $\nu$ and $\bar{\nu}$ modes. 
The MH sensitivity of DUNE using only the standard LE beam is already very high for the entire range of values of the CP phase and $\area{MH}$ combination does not offer much improvement. Since the contribution to MH sensitivity mainly comes from around the first oscillation maximum ($2-2.5$ GeV) for which the LE beam offers best statistics, the use of ME beam is not expected to lead to much improvement.  The best MH runtime combination is (${\bf{4.5+1.5+0.5+0.5}}$) for NH  and ($2.5+4+0.5+0$) for IH. For this particular question also,  the $\numu$ and $\nutau$ channels do not contribute significantly (see Appendix~{\bf A} and  \cite{Masud:2016gcl} to get an insight into the typical shape of the MH sensitivities at long baseline experiments as well as for an understanding of  the role of different oscillation channels). As can be seen from Fig.~\ref{fig:opt_mh}, we have also estimated the MH sensitivity corresponding to $\area{CPV}$ (brown dashed curve). It is seen that for NH, it  is almost the same as that of $\area{MH}$ while for IH it is lower than the standard sensitivity (black curve). Interestingly, we observe that, for $\area{MH}$, $\len$ plays a dominant role when the hierarchy is normal while $\lenbar$ plays a dominant role for the IH case.  
%%%%%%%%%%%%%%%%%%%%%%%%
\begin{figure}[htb]
\centering
\includegraphics[scale=0.5]{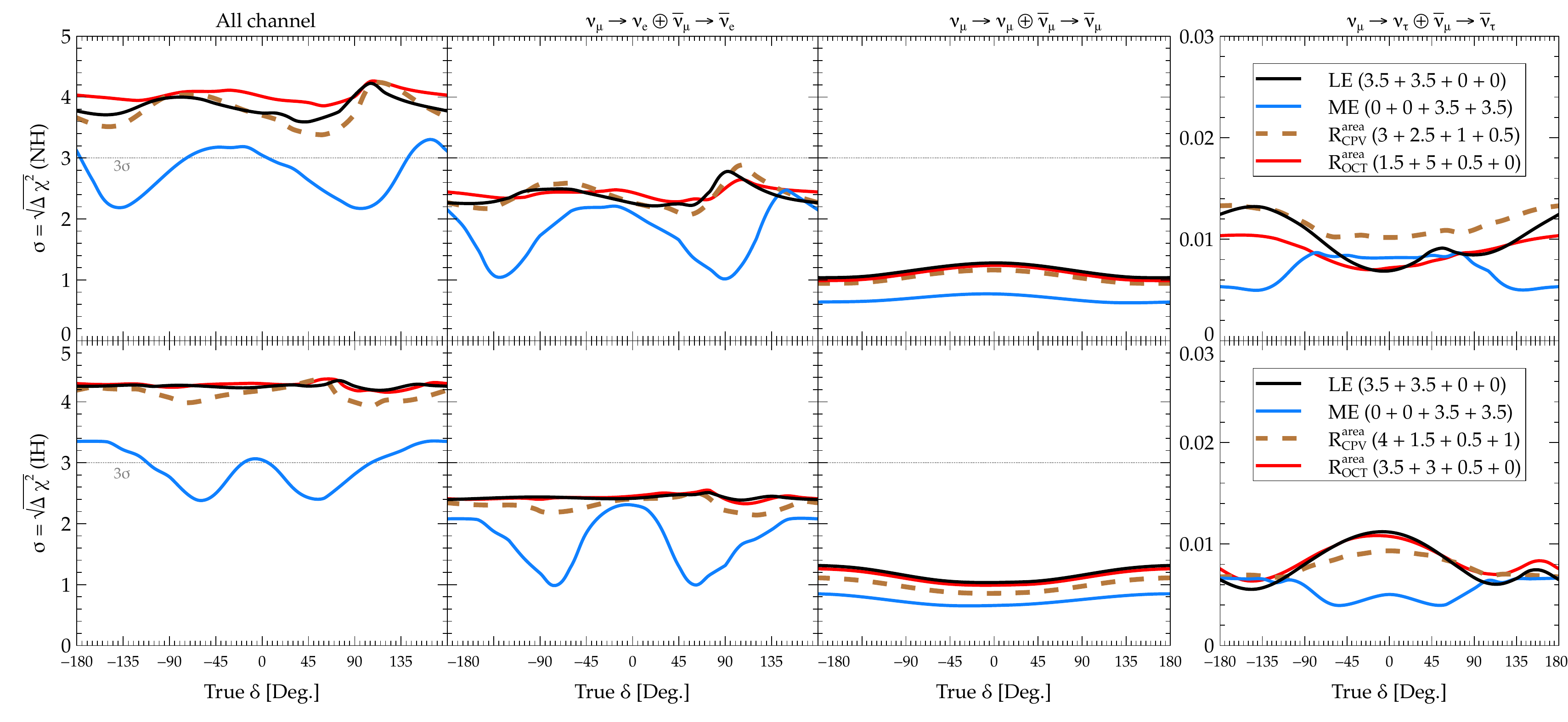}
\caption{\footnotesize{The sensitivities to $\tc$-octant degeneracy (true octant being HO) at DUNE are shown with the variation of true $\delta$, for optimized runtime combination for LE+ME, - found by maximizing the area under the sensitivity curve (solid red). 
The sensitivities obtained for LE only(black) or ME only (blue) beam are also shown with equal runtime in $\nu$ and $\bar{\nu}$ modes. 
The legends signify the runtime combination for each case in the form ($\len + \lenbar + \men + \menbar$). 
For comparison, the octant sensitivities corresponding to $\area{CPV}$ are also shown (dashed brown). 
The legends signify the runtime combination for each case in the form ($\len + \lenbar + \men + \menbar$). 
The top (bottom) row depicts the case of true NH (IH).
The first column shows the combined results considering all channels while the second, third and fourth column shows the contributions from the individual channels $\nue, \numu$ and $\nutau$ respectively.}}
\label{fig:opt_octant_ho}
\end{figure}
%%%%%%%%%%%%%%%%%%%%%%%

{\underline{\bf{Sensitivity to octant of $\tc$:-}}}
We next compute the optimized runtime combinations of $(\len + \lenbar + \men + \menbar)$ that gives the best sensitivity to $\tc$ octant at DUNE when the true octant is HO ($\tc$ = $48.8^{\circ}$). In Fig.\ \ref{fig:opt_octant_ho}, the top (bottom) row depict the case of NH (IH), while the second, third and fourth  columns show the roles of individual oscillation channels. 
We observe in Fig.\ \ref{fig:opt_octant_ho} that the best runtime combination of (${\bf{1.5 + 5 + 0.5 + 0}}$) for NH scenario underscores the importance of $\bar{\nu}$ runtimes\footnote{As has been discussed in detail in \cite{Ghosh:2014rna}, antineutrino run is important in estimating the $\tc$ octant degeneracy due to complementary nature of the degeneracy for $\nue$ and $\nuebar$ channels.} and this optimized combination improves this sensitivity significantly. 
This is apparent for IH scenario ($2.5 + 3 + 1.5 + 0$) as well, albeit in a less prominent manner. 
We also note that unlike the case of CPV and MH, the $\numu$ ($\numubar$) disappearance channel contributes substantially to the octant sensitivities. 
The $\nutau$ ($\nutaubar$) channel, on the other hand, practically has no role (see Appendix~{\bf A}).   
We also show the octant sensitivities by using only the standard LE beam (black) or the ME beam only (blue), the distribution of runtime being equally  shared among the $\nu$ and $\bar{\nu}$ modes. 
Additionally, for the sake of comparison, the dashed brown curves illustrate the octant sensitivity obtained corresponding to $\area{CPV}$. 

Tab.\ \ref{tab:opt} summarises the results for estimated optimized combinations with respect to CPV sensitivity, MH sensitivity and sensitivity to the octant of $\theta_{23}$ respectively.  
%%%%%%%%%%%%%%%%%%%%%%%%%%
\begin{table}[ht]
\centering
\begin{tabular}{|c|c|c|c|}
 \hline
  Sensitivity to & Optimization combination & NH  & IH\\ 
  && $(\len + \lenbar + \men + \menbar)$ & $(\len + \lenbar + \men + \menbar)$\\
   && (in years)  & (in years) \\   \hline
   \multirow{2}{*}{CPV} & $\area{CPV}$ & $3 + 2.5 + 1 + 0.5$ & $4 + 1.5 + 0.5 + 1$\\ \cline{2-4}
   & $\fraction{CPV}$ & $3 + 2 + 1 + 1$ & $4 + 1.5 + 1 + 0.5$\\ \hline
   MH & $\area{MH}$ & $4.5 + 1.5 + 0.5 + 0.5$ & $2.5 + 4 + 0.5 + 0$\\ \hline
   Octant of $\tc$ & $\area{OCT}$ (HO) & $1.5 + 5 + 0.5 + 0$ & $3.5 + 3 + 0.5 + 0$\\ \hline
%   \multirow{2}{*}{Octant of $\tc$} & $\area{OCT}$ (HO) & $1.5 + 5 + 0.5 + 0$ & $3.5 + 3 + 0.5 + 0$\\ \cline{2-4}
%   & $\area{OCT}$ (LO) & $3 + 3 + 1 + 0$ & $1.5 + 4.5 + 1 + 0$\\ \hline
\end{tabular}
\caption{Optimized runtime combinations for sensitivity to CPV, MH and   octant  of $\tc$  at DUNE.}
\label{tab:opt}
\end{table}
%%%%%%%%%%%%%%%%%%%%%%%%%%
%===================================
\section{Sensitivity to the choice of optimal combination}
\label{sec:heatmap}
%=======================================
Having completed the task of estimating the optimal combination of runtime that yields the best sensitivity to the three unknowns, we would now like to pose the following question. How sensitive are we to the choice of optimal combination or what direction can we take (in our choice of runtime combinations) in case we have difficulty in implementation of the particular runtime combination. We address these questions in the present section. 

\begin{figure}[hbt!]
\centering
\includegraphics[scale=0.5]{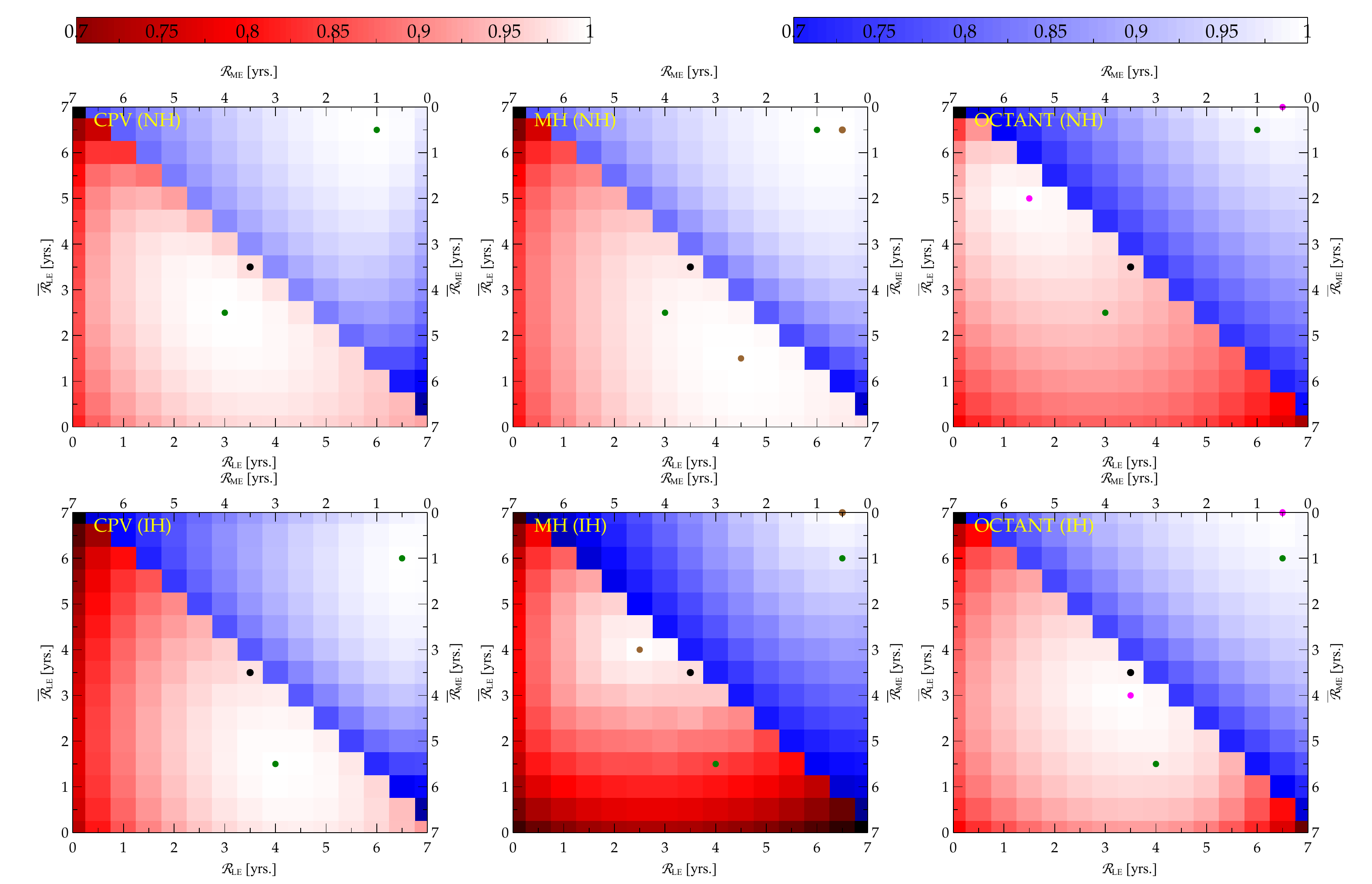}
\caption{\footnotesize{Heatmap of the normalized area under the sensitivity curves (in the [$\chisq$-true $\delta$] plane) for all the runtime combinations considered in the context of CPV, MH and $\tc$ octant sensitivities in the three columns respectively. 
The top (bottom) row depicts the case of true NH (IH). 
The four components of the runtime combination are shown along the four sides of each panel.
The lighter (darker) shades of the colours imply better (worse) sensitivity. 
We mark the case of standard runtime ($\len + \lenbar + \men + \menbar = 3.5 + 3.5 + 0 + 0$) with a black dot. 
The optimized runtime combinations summarised in Tab.\ \ref{tab:opt} ($\area{CPV}, \area{MH}, \area{OCT}$) giving 
the best sensitivities are marked with green, brown and magenta dots in the three columns respectively. 
The combination $\area{CPV}$ (green dot) is also marked for 
the case of MH and octant sensitivities (\ie, second and third column) for comparison.
}}
\label{fig:visualize}
\end{figure}

We examine the different runtime combinations and analyze how the results improve (\ie, how the area under the $\chisq$ curve as a function of true $\delta$ for CPV, MH, octant increases) for different  runtime combinations.
In Fig.\ \ref{fig:visualize}, we show the heatmap of the area (normalized) under the 
 sensitivity curves (in the $\chisq$-true $\delta$ plane) for all the runtime combinations ($\len + \lenbar + \men + \menbar $) considered. 
 The three columns show the case of CPV, MH and octant sensitivities, while the top (bottom) 
 row depicts the NH (IH) scenario.
The lower triangular portion (red) shows the effect of the $\len$ and $\lenbar$ components of the runtime combination along the bottom horizontal axis and the left vertical axis respectively.
The upper triangular portion (blue) shows the effect of the $\men$ and $\menbar$ components of the runtime combination along the top horizontal axis and the right vertical axis respectively.  The lighter (darker) shades of the colours imply better (worse) sensitivity. 
Thus, any runtime combination read off Fig.\ \ref{fig:visualize} consists of two parts: ($\len + \lenbar$) in the red lower triangular region, and the corresponding ($\men + \menbar$) in the blue upper triangular region such that $\len + \lenbar + \men + \menbar = 7$. 
The black dot indicates the case of standard runtime using only LE beam (${\bf{3.5 + 3.5 + 0 + 0}}$).  
The optimized runtime combinations ($\area{CPV}, \area{MH}, \area{OCT}$) giving 
the best sensitivities are marked with a pair of green, brown and magenta dots in the three columns respectively. 
For comparison, the combination $\area{CPV}$ (green dot) is also marked for 
the case of MH and octant sensitivities (\ie, second and third column).
The fact that the lighter shaded regions ae located away from the combination represented by the black dot clearly indicates how the result improves when one combines the ME beam (\ie, nonzero $\men$ and $\menbar$) with the LE beam.
For MH (2nd column), interestingly more $\len$ facilitates the improvement of the sensitivities for true NH case, while $\lenbar$ is slightly favoured for true IH. 
For octant sensitivity (third column of Fig.\ \ref{fig:visualize}) it is clear that for NH, $\area{OCT}$ is dominated by $\lenbar$, while for IH, both $\len$ and $\lenbar$ play equally important roles. 
Fig.\ \ref{fig:visualize} also helps to highlight the need for $\bar{\nu}$ mode runs in probing the $\tc$ octant sensitivity. 
%%%%%%%%%%%%%%%%%%%%%%%%
%%%%%%%%%%%%%%%%%%%%%%%%%%%%
%=================================
\section{Summary}
\label{sec:summary}
%==================================
CP violation, MH and octant of $\theta_{23}$ are the crucial unknowns and current and future long baseline experiments such as DUNE are planned to address these questions. In the basic configuration, it is assumed that DUNE would have a  runtime of 7 years (distributed equally in the $\nu$ and $\bar{\nu}$ mode) with the standard LE beam.  The  LE beam that is often used in DUNE simulations has a peak around $2-3$ GeV (the first oscillation maximum for $\nue$ transition probability) but very sharply falls off at $E \gtrsim 4$ GeV.  Consequently, the number of events beyond $4$ GeV rapidly becomes smaller, providing very little statistics.  In the present work, we propose to use a higher energy, ME beam that has a substantial flux even beyond 4 GeV in addition to the LE beam and ask whether this can offer any improvement to the standard sensitivity reach of DUNE in answering question pertaining to CPV, MH and octant of $\tc$.

Keeping the total runtime fixed to 7 years, we have distributed the total runtime among the $\nu$ and $\bar{\nu}$ modes with the possibility of  
utilizing the different beam tunes, LE and ME.  {\hlpm{We have chosen a design for the ME beam that is nominally compatible with the space and infrastructure capabilities of the LBNF/DUNE beamline and that is based on a real working beamline design currently deployed in NuMI/NOvA.}}  In each of these sensitivity analyses, we have considered the (neutrino and antineutrino) contributions of all three oscillation channels $\nue, \numu$ and $\nutau$ and shown   the contribition of the individual channels to the overall sensitivity. We  specify the different runtimes using different beam tunes and modes as $\len, \lenbar, \men, \menbar$.  The  optimized combinations $\len + \lenbar + \men + \menbar$  that give the best sensitivities to CPV, MH and octant of $\tc$ are then evaluated.

Our results are reported  in Sec.~\ref{sec:opt}. In Fig.~\ref{fig:opt_cp}, we found that a runtime combination of ${\bf{(3 + 2.5 + 1 + 0.5)}}$   gives the best sensitivity to CP violation if the hierarchy is normal. Also, the sensitivity can reach beyond $5\sigma$, which was otherwise not possible with the standard DUNE configuration with LE beam alone near $\delta \simeq \pm \pi/2$ (maximal CPV). In addition to resolving CPV, this particular optimized runtime combination also offers high sensitivity to resolve the MH (see Fig.~\ref{fig:opt_mh}) and octant of $\tc$ (see Fig.~\ref{fig:opt_octant_ho}). For MH and octant of $\tc$, the optimized runtime combinations providing the best sensitivities are found to be ${\bf{(4.5 + 1.5 + 0.5 + 0.5)}}$ and ${\bf{(1.5 + 5 + 0.5 + 0)}}$ respectively (assuming the hierarchy is normal). Finally, Tab.\ \ref{tab:opt} summarises the results for estimated optimized combinations w.r.t. CPV sensitivity, MH sensitivity and senstivity to the octant of $\theta_{23}$ respectively.  
 
This study, therefore, underscores the availability of the room for improvement within the DUNE experimental configuration by using a combination of runtime in the $\nu$ and $\bar{\nu}$ mode, exploiting two different (LE and ME) beam tunes. This suggested runtime configuration with the two available beam tunes will eventually help DUNE to answer, with  more  robustness, its main goals pertaining to leptonic CP violation, determination of MH and octant of $\tc$.  

{\hlpm{We would like to mention that our phenomenological study is concerned primarily with bringing out the physics capabilities introduced by the ME beam tune. Consideration of the cost as well as other detailed technical changes needed to replace the current beamline design with the ME tune proposed above is outside the purview of this paper. It may be worth mentioning that LBNF/DUNE full CP violation sensitivity assumes an upgrade to a 2.4 MW beam which will require a complete redesign of the targeting and focusing system and thus the facility is designed with the ability to accommodate different targetry and focusing designs.}}

%==========================
\acknowledgements 
M.M. acknowledges the financial support from the Indian National Science Academy (INSA) Young Scientist Project  [INSA/SP/YS/2019/269]. The authors acknowledge HPC Cluster Computing System at HRI. 
This material is based upon work supported by the U.S. Department of Energy, Office of Science, Office of High Energy Physics under contract number DE-SC0012704; the Indian funding from University Grants Commission under the second phase of University with Potential of Excellence (UPE II) at JNU and Department of Science and Technology under DST-PURSE at JNU. This work has received partial funding from the European Union’s Horizon 2020 research and innovation programme under the Marie Skodowska-Curie grant agreement No 690575 and 674896. PM acknowledges the kind hospitality from the particle physics group at Brookhaven National Laboratory during the initial stages of  this work. 

%===================================
\appendix
\renewcommand\theequation{A\arabic{equation}}
\setcounter{equation}{0} 
\section*{Appendix A: Role of different channels in probability}
\label{appendix}
%===================================
%%%%%%%%%%%%%%%%%%%%
\begin{figure}[th!]
\centering
\includegraphics[scale=0.5]
{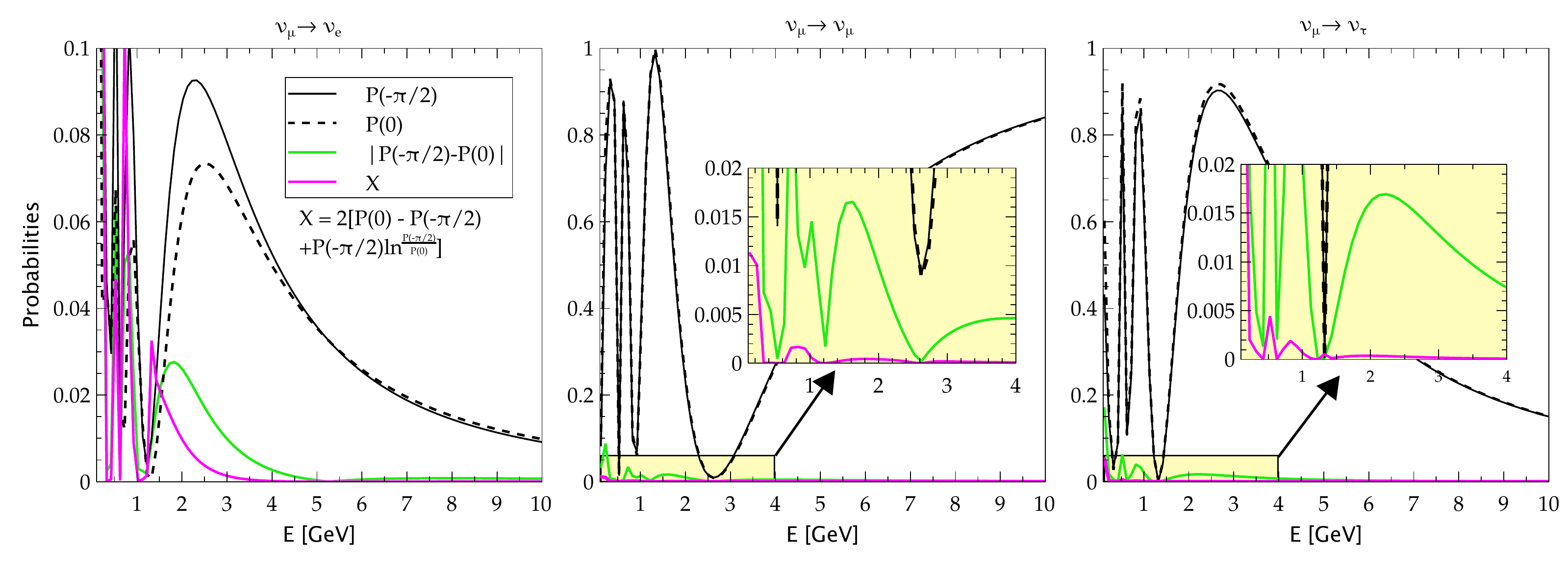}
\caption{\footnotesize{Comparison of probabilities (solid and dashed black) and their differences (green) for the three channels $\nue$ (first panel), $\numu$ (second) and $\nutau$ (third). 
A $\chisq$-like quantity $X$ (magenta) in the probability level gives a measure of the contributions of these three channels to the sensitivity to CPV.
}}
\label{fig:prob_cp}
\end{figure}
%%%%%%%%%%%%%%%%%
Here we discuss how the individual channels ($\nue, \numu, \nutau$) contribute in the probability level in probing the questions related  
 to CPV, MH and the octant of $\tc$ at the DUNE baseline of 1300 km. 
 The $\chisq$ estimation for CPV in Fig.\ \ref{fig:opt_cp} gives a numerical measure of the {\it{difference}} between the CP conserving value (test $\delta = 0$ or $\pi$) and all values of true $\delta$ ($\in [-\pi, \pi]$) and is maximum around true $\delta \approx \pm \pi/2$. 
 In Fig.\ \ref{fig:prob_cp}, we do a probability analysis by plotting $P_{\mu\beta}$ ($\beta = e, \mu, \tau$) for $\delta = -\pi/2$ (solid black), $0$ (dashed black) and their absolute difference (green). 
 Finally we also show a $\chisq$-like quantity $X$ (magenta) defined in the probability level following the statistical part of Eq.\ \ref{eq:chisq}. 
 For Fig.\ \ref{fig:prob_cp}, the explicit definition of this quantity is,
 \begin{equation}
 \label{eq:X_cp}
 X_{\text{CPV}} = 2\bigg[P_{\mu\beta}(0) - P_{\mu\beta}(-\pi/2) + P_{\mu\beta}(-\pi/2) \ln \frac{P_{\mu\beta}(-\pi/2)}{P_{\mu\beta}(0)}
 \bigg],
 \end{equation}
 where the argument within the parentheses are the values of the CP phase $\delta$.
  Fig.\ \ref{fig:prob_cp} shows that though the magnitudes of the difference of probabilities (green) are in the similar ballpark for all the three channels, the contribution to the $\chisq$-like quantity $X$ (magenta) mainly comes from the $\nue$ channel. 
  This is because the fractional difference of the probabilities (which dominates the estimation of $\chisq$) in the $\nue$ channel is much higher,- owing to the small magnitudes of $P_{\mu e}$. 
  As is clear from the insets in Fig.\ \ref{fig:prob_cp}, this fractional difference (magenta) is tiny for the $\numu$ and $\nutau$ channels. 

%%%%%%%%
\begin{figure}[th!]
\centering
\includegraphics[scale=0.5]
{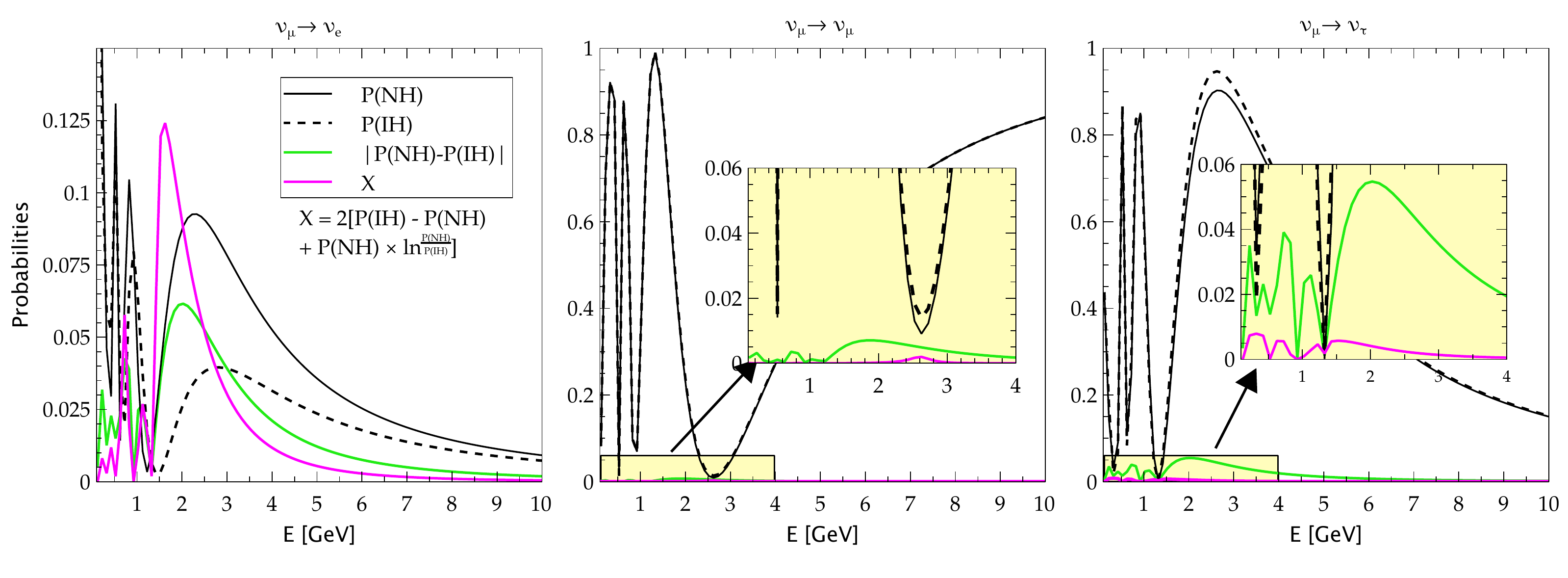}
\caption{\footnotesize{
Similar to Fig.\ \ref{fig:prob_cp} but for contribution to explore the MH degeneracy.
}}
\label{fig:prob_mh}
\end{figure}
%%%%% 
 Similarly, in Fig.\ \ref{fig:prob_mh}, we analyse the contributions of the three channels in probing the MH degeneracy. 
 For the two opposite hierarchies NH and IH, we show $P_{\mu\beta}$ (solid black and dashed black), their difference (green) and the $\chisq$-like quantity (magenta) defined below:
  \begin{equation}
 \label{eq:X_mh}
 X_{\text{MH}} = 2\bigg[P_{\mu\beta}(IH) - P_{\mu\beta}(NH) + P_{\mu\beta}(NH) \ln \frac{P_{\mu\beta}(NH)}{P_{\mu\beta}(IH)}
 \bigg]. 
 \end{equation} 
We take the CP phase to be $-\pi/2$ here. 
It can be easily observed that $X$ is again dominated by the $\nue$ channel, while the $\numu$ channel gives almost no contribution. 
Interestingly, though the difference of $P_{\mu\tau}$ for NH and IH (\ie, green curve) is very similar in magnitude with that of $P_{\mu e}$, the large value of $P_{\mu\tau}$ makes the fractional difference and consequently the value of the $\chisq$-like quantity $X$ insignificant.

\begin{figure}[th!]
\centering
\includegraphics[scale=0.5]
{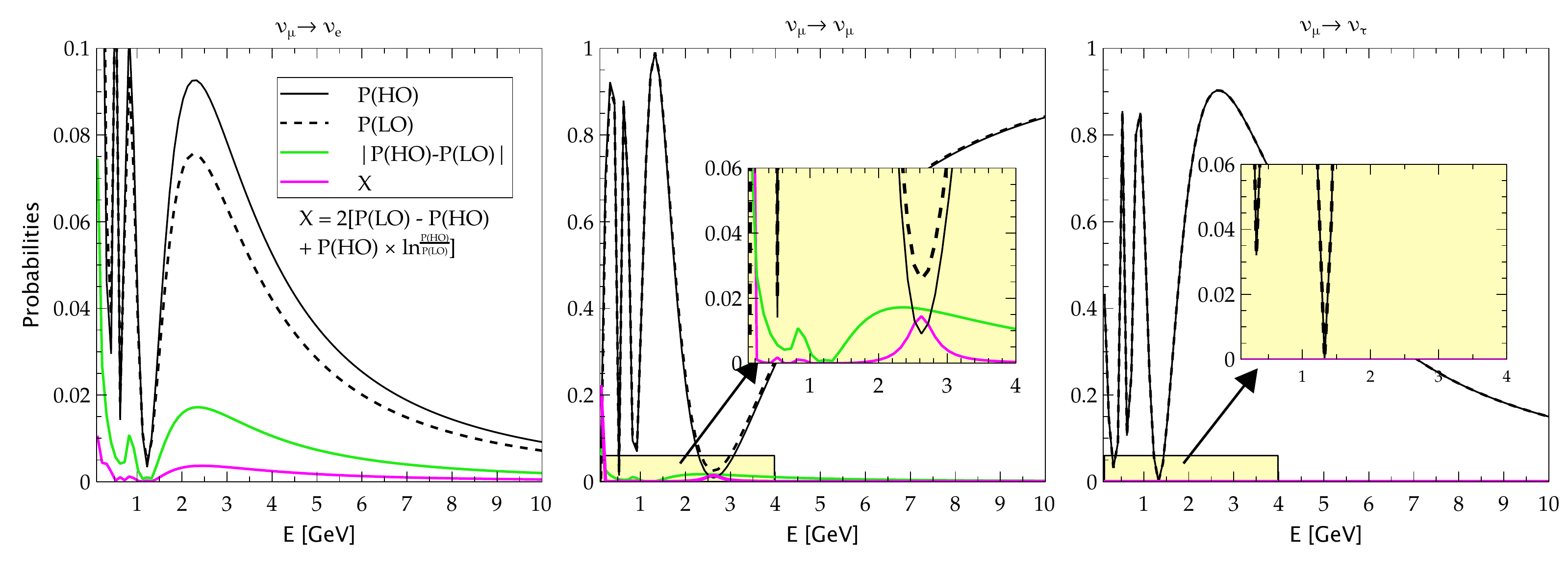}
\caption{\footnotesize{Similar to Fig.\ \ref{fig:prob_cp} but for contribution to explore the $\tc$-octant degeneracy.
}}
\label{fig:prob_octant}
\end{figure}

Finally the probability level analyses for $\tc$-octant degeneracy is illustrated in Fig.\ \ref{fig:prob_octant}. 
The $\chisq$-like quantity $X$ is defined as follows.  
  \begin{equation}
 \label{eq:X_octant}
 X_{\text{OCT}} = 2\bigg[P_{\mu\beta}(LO) - P_{\mu\beta}(HO) + P_{\mu\beta}(HO) \ln \frac{P_{\mu\beta}(HO)}{P_{\mu\beta}(LO)}
 \bigg], 
 \end{equation} 
 where the value of the CP phase $\delta$ was kept at its best fit value of $-\pi/2$.
We see both the $\nue$ and $\numu$ channel contribute to the octant sensitivity, the latter slightly dominating around the crucial energy region of $2-3$ GeV\footnote{The very small magnitude of $P_{\mu\mu}$ around this energy range helps to enhance $\chisq$-like quantity X.}.   
%%%%%%%%%%%%%%%%%%%%%
 %%%%%%%%%%%%%%%%%%%%%
\bibliographystyle{apsrev}
\bibliography{referencesnsi}

\begin{thebibliography}{65}
\expandafter\ifx\csname natexlab\endcsname\relax\def\natexlab#1{#1}\fi
\expandafter\ifx\csname bibnamefont\endcsname\relax
  \def\bibnamefont#1{#1}\fi
\expandafter\ifx\csname bibfnamefont\endcsname\relax
  \def\bibfnamefont#1{#1}\fi
\expandafter\ifx\csname citenamefont\endcsname\relax
  \def\citenamefont#1{#1}\fi
\expandafter\ifx\csname url\endcsname\relax
  \def\url#1{\texttt{#1}}\fi
\expandafter\ifx\csname urlprefix\endcsname\relax\def\urlprefix{URL }\fi
\providecommand{\bibinfo}[2]{#2}
\providecommand{\eprint}[2][]{\url{#2}}

\bibitem[{\citenamefont{Cowan et~al.}(1956)\citenamefont{Cowan, Reines,
  Harrison, Kruse, and McGuire}}]{Cowan:1992xc}
\bibinfo{author}{\bibfnamefont{C.}~\bibnamefont{Cowan}},
  \bibinfo{author}{\bibfnamefont{F.}~\bibnamefont{Reines}},
  \bibinfo{author}{\bibfnamefont{F.}~\bibnamefont{Harrison}},
  \bibinfo{author}{\bibfnamefont{H.}~\bibnamefont{Kruse}}, \bibnamefont{and}
  \bibinfo{author}{\bibfnamefont{A.}~\bibnamefont{McGuire}},
  \bibinfo{journal}{Science} \textbf{\bibinfo{volume}{124}},
  \bibinfo{pages}{103} (\bibinfo{year}{1956}).

\bibitem[{\citenamefont{Pontecorvo}(1957)}]{Pontecorvo:1957cp}
\bibinfo{author}{\bibfnamefont{B.}~\bibnamefont{Pontecorvo}},
  \bibinfo{journal}{Sov. Phys. JETP} \textbf{\bibinfo{volume}{6}},
  \bibinfo{pages}{429} (\bibinfo{year}{1957}).

\bibitem[{\citenamefont{Pontecorvo}(1958)}]{Pontecorvo:1957qd}
\bibinfo{author}{\bibfnamefont{B.}~\bibnamefont{Pontecorvo}},
  \bibinfo{journal}{Sov. Phys. JETP} \textbf{\bibinfo{volume}{7}},
  \bibinfo{pages}{172} (\bibinfo{year}{1958}).

\bibitem[{\citenamefont{Maki et~al.}(1962)\citenamefont{Maki, Nakagawa, and
  Sakata}}]{mns}
\bibinfo{author}{\bibfnamefont{Z.}~\bibnamefont{Maki}},
  \bibinfo{author}{\bibfnamefont{M.}~\bibnamefont{Nakagawa}}, \bibnamefont{and}
  \bibinfo{author}{\bibfnamefont{S.}~\bibnamefont{Sakata}},
  \bibinfo{journal}{Progress of Theoretical Physics}
  \textbf{\bibinfo{volume}{28}}, \bibinfo{pages}{870} (\bibinfo{year}{1962}).

\bibitem[{\citenamefont{Gribov and Pontecorvo}(1969)}]{Gribov:1968kq}
\bibinfo{author}{\bibfnamefont{V.}~\bibnamefont{Gribov}} \bibnamefont{and}
  \bibinfo{author}{\bibfnamefont{B.}~\bibnamefont{Pontecorvo}},
  \bibinfo{journal}{Phys. Lett. B} \textbf{\bibinfo{volume}{28}},
  \bibinfo{pages}{493} (\bibinfo{year}{1969}).

\bibitem[{\citenamefont{Kajita and McDonald}()}]{nobel2015}
\bibinfo{author}{\bibfnamefont{T.}~\bibnamefont{Kajita}} \bibnamefont{and}
  \bibinfo{author}{\bibfnamefont{A.~B.} \bibnamefont{McDonald}},
  \bibinfo{note}{the Nobel Prize in Physics
  2015.\url{https://www.nobelprize.org/prizes/physics/2015/summary/}}.

\bibitem[{\citenamefont{Zyla et~al.}(2020)}]{Zyla:2020zbs}
\bibinfo{author}{\bibfnamefont{P.}~\bibnamefont{Zyla}} \bibnamefont{et~al.}
  (\bibinfo{collaboration}{Particle Data Group}), \bibinfo{journal}{PTEP}
  \textbf{\bibinfo{volume}{2020}}, \bibinfo{pages}{083C01}
  (\bibinfo{year}{2020}).

\bibitem[{\citenamefont{Sakharov}(1991)}]{Sakharov:1967dj}
\bibinfo{author}{\bibfnamefont{A.}~\bibnamefont{Sakharov}},
  \bibinfo{journal}{Sov. Phys. Usp.} \textbf{\bibinfo{volume}{34}},
  \bibinfo{pages}{392} (\bibinfo{year}{1991}).

\bibitem[{\citenamefont{Fukugita and Yanagida}(1986)}]{Fukugita:1986hr}
\bibinfo{author}{\bibfnamefont{M.}~\bibnamefont{Fukugita}} \bibnamefont{and}
  \bibinfo{author}{\bibfnamefont{T.}~\bibnamefont{Yanagida}},
  \bibinfo{journal}{Phys. Lett.} \textbf{\bibinfo{volume}{B174}},
  \bibinfo{pages}{45} (\bibinfo{year}{1986}).

\bibitem[{\citenamefont{Davidson et~al.}(2008)\citenamefont{Davidson, Nardi,
  and Nir}}]{Davidson:2008bu}
\bibinfo{author}{\bibfnamefont{S.}~\bibnamefont{Davidson}},
  \bibinfo{author}{\bibfnamefont{E.}~\bibnamefont{Nardi}}, \bibnamefont{and}
  \bibinfo{author}{\bibfnamefont{Y.}~\bibnamefont{Nir}},
  \bibinfo{journal}{Phys. Rept.} \textbf{\bibinfo{volume}{466}},
  \bibinfo{pages}{105} (\bibinfo{year}{2008}), \eprint{0802.2962}.

\bibitem[{\citenamefont{Branco et~al.}(2012)\citenamefont{Branco, Felipe, and
  Joaquim}}]{Branco:2011zb}
\bibinfo{author}{\bibfnamefont{G.~C.} \bibnamefont{Branco}},
  \bibinfo{author}{\bibfnamefont{R.~G.} \bibnamefont{Felipe}},
  \bibnamefont{and} \bibinfo{author}{\bibfnamefont{F.~R.}
  \bibnamefont{Joaquim}}, \bibinfo{journal}{Rev. Mod. Phys.}
  \textbf{\bibinfo{volume}{84}}, \bibinfo{pages}{515} (\bibinfo{year}{2012}),
  \eprint{1111.5332}.

\bibitem[{\citenamefont{Petcov}(1982)}]{Petcov:1982ya}
\bibinfo{author}{\bibfnamefont{S.}~\bibnamefont{Petcov}},
  \bibinfo{journal}{Phys. Lett. B} \textbf{\bibinfo{volume}{110}},
  \bibinfo{pages}{245} (\bibinfo{year}{1982}).

\bibitem[{\citenamefont{King et~al.}(2014)\citenamefont{King, Merle, Morisi,
  Shimizu, and Tanimoto}}]{King:2014nza}
\bibinfo{author}{\bibfnamefont{S.~F.} \bibnamefont{King}},
  \bibinfo{author}{\bibfnamefont{A.}~\bibnamefont{Merle}},
  \bibinfo{author}{\bibfnamefont{S.}~\bibnamefont{Morisi}},
  \bibinfo{author}{\bibfnamefont{Y.}~\bibnamefont{Shimizu}}, \bibnamefont{and}
  \bibinfo{author}{\bibfnamefont{M.}~\bibnamefont{Tanimoto}},
  \bibinfo{journal}{New J. Phys.} \textbf{\bibinfo{volume}{16}},
  \bibinfo{pages}{045018} (\bibinfo{year}{2014}), \eprint{1402.4271}.

\bibitem[{\citenamefont{Haxton and Stephenson}(1984)}]{Haxton:1985am}
\bibinfo{author}{\bibfnamefont{W.}~\bibnamefont{Haxton}} \bibnamefont{and}
  \bibinfo{author}{\bibfnamefont{G.}~\bibnamefont{Stephenson}},
  \bibinfo{journal}{Prog. Part. Nucl. Phys.} \textbf{\bibinfo{volume}{12}},
  \bibinfo{pages}{409} (\bibinfo{year}{1984}).

\bibitem[{\citenamefont{Elliott and Engel}(2004)}]{Elliott:2004hr}
\bibinfo{author}{\bibfnamefont{S.~R.} \bibnamefont{Elliott}} \bibnamefont{and}
  \bibinfo{author}{\bibfnamefont{J.}~\bibnamefont{Engel}}, \bibinfo{journal}{J.
  Phys. G} \textbf{\bibinfo{volume}{30}}, \bibinfo{pages}{R183}
  (\bibinfo{year}{2004}), \eprint{hep-ph/0405078}.

\bibitem[{\citenamefont{Aalseth et~al.}(2004)}]{Aalseth:2004hb}
\bibinfo{author}{\bibfnamefont{C.}~\bibnamefont{Aalseth}} \bibnamefont{et~al.}
  (\bibinfo{year}{2004}), \eprint{hep-ph/0412300}.

\bibitem[{\citenamefont{Lam}(2001)}]{Lam:2001fb}
\bibinfo{author}{\bibfnamefont{C.}~\bibnamefont{Lam}}, \bibinfo{journal}{Phys.
  Lett. B} \textbf{\bibinfo{volume}{507}}, \bibinfo{pages}{214}
  (\bibinfo{year}{2001}), \eprint{hep-ph/0104116}.

\bibitem[{\citenamefont{Harrison and Scott}(2002)}]{Harrison:2002et}
\bibinfo{author}{\bibfnamefont{P.}~\bibnamefont{Harrison}} \bibnamefont{and}
  \bibinfo{author}{\bibfnamefont{W.}~\bibnamefont{Scott}},
  \bibinfo{journal}{Phys. Lett. B} \textbf{\bibinfo{volume}{547}},
  \bibinfo{pages}{219} (\bibinfo{year}{2002}), \eprint{hep-ph/0210197}.

\bibitem[{\citenamefont{Xing and Zhao}(2016)}]{Xing:2015fdg}
\bibinfo{author}{\bibfnamefont{Z.-z.} \bibnamefont{Xing}} \bibnamefont{and}
  \bibinfo{author}{\bibfnamefont{Z.-h.} \bibnamefont{Zhao}},
  \bibinfo{journal}{Rept. Prog. Phys.} \textbf{\bibinfo{volume}{79}},
  \bibinfo{pages}{076201} (\bibinfo{year}{2016}), \eprint{1512.04207}.

\bibitem[{\citenamefont{Abe et~al.}(2014)}]{Abe:2013hdq}
\bibinfo{author}{\bibfnamefont{K.}~\bibnamefont{Abe}} \bibnamefont{et~al.}
  (\bibinfo{collaboration}{T2K}), \bibinfo{journal}{Phys. Rev. Lett.}
  \textbf{\bibinfo{volume}{112}}, \bibinfo{pages}{061802}
  (\bibinfo{year}{2014}), \eprint{1311.4750}.

\bibitem[{\citenamefont{Ayres et~al.}(2004)}]{Ayres:2004js}
\bibinfo{author}{\bibfnamefont{D.~S.} \bibnamefont{Ayres}} \bibnamefont{et~al.}
  (\bibinfo{collaboration}{NOvA}) (\bibinfo{year}{2004}),
  \eprint{hep-ex/0503053}.

\bibitem[{\citenamefont{Abe et~al.}(2020)}]{Abe:2019vii}
\bibinfo{author}{\bibfnamefont{K.}~\bibnamefont{Abe}} \bibnamefont{et~al.}
  (\bibinfo{collaboration}{T2K}), \bibinfo{journal}{Nature}
  \textbf{\bibinfo{volume}{580}}, \bibinfo{pages}{339} (\bibinfo{year}{2020}),
  \bibinfo{note}{[Erratum: Nature 583, E16 (2020)]}, \eprint{1910.03887}.

\bibitem[{\citenamefont{Acero et~al.}(2019)}]{Acero:2019ksn}
\bibinfo{author}{\bibfnamefont{M.~A.} \bibnamefont{Acero}} \bibnamefont{et~al.}
  (\bibinfo{collaboration}{NOvA}), \bibinfo{journal}{Phys. Rev. Lett.}
  \textbf{\bibinfo{volume}{123}}, \bibinfo{pages}{151803}
  (\bibinfo{year}{2019}), \eprint{1906.04907}.

\bibitem[{\citenamefont{de~Salas et~al.}(2020)\citenamefont{de~Salas, Forero,
  Gariazzo, Mart{\'i}nez-Mirav{\'e}, Mena, Ternes, T{\'o}rtola, and
  Valle}}]{deSalas:2020pgw}
\bibinfo{author}{\bibfnamefont{P.}~\bibnamefont{de~Salas}},
  \bibinfo{author}{\bibfnamefont{D.}~\bibnamefont{Forero}},
  \bibinfo{author}{\bibfnamefont{S.}~\bibnamefont{Gariazzo}},
  \bibinfo{author}{\bibfnamefont{P.}~\bibnamefont{Mart{\'i}nez-Mirav{\'e}}},
  \bibinfo{author}{\bibfnamefont{O.}~\bibnamefont{Mena}},
  \bibinfo{author}{\bibfnamefont{C.}~\bibnamefont{Ternes}},
  \bibinfo{author}{\bibfnamefont{M.}~\bibnamefont{T{\'o}rtola}},
  \bibnamefont{and} \bibinfo{author}{\bibfnamefont{J.}~\bibnamefont{Valle}}
  (\bibinfo{year}{2020}), \eprint{2006.11237}.

\bibitem[{\citenamefont{Valencia-Globalfit}(2020)}]{globalfit}
\bibinfo{author}{\bibnamefont{Valencia-Globalfit}},
  \bibinfo{howpublished}{\url{http://globalfit.astroparticles.es/}}
  (\bibinfo{year}{2020}).

\bibitem[{\citenamefont{Capozzi et~al.}(2018)\citenamefont{Capozzi, Lisi,
  Marrone, and Palazzo}}]{Capozzi:2018ubv}
\bibinfo{author}{\bibfnamefont{F.}~\bibnamefont{Capozzi}},
  \bibinfo{author}{\bibfnamefont{E.}~\bibnamefont{Lisi}},
  \bibinfo{author}{\bibfnamefont{A.}~\bibnamefont{Marrone}}, \bibnamefont{and}
  \bibinfo{author}{\bibfnamefont{A.}~\bibnamefont{Palazzo}},
  \bibinfo{journal}{Prog. Part. Nucl. Phys.} \textbf{\bibinfo{volume}{102}},
  \bibinfo{pages}{48} (\bibinfo{year}{2018}), \eprint{1804.09678}.

\bibitem[{\citenamefont{Esteban et~al.}(2019)\citenamefont{Esteban,
  Gonzalez-Garcia, Hernandez-Cabezudo, Maltoni, and Schwetz}}]{Esteban:2018azc}
\bibinfo{author}{\bibfnamefont{I.}~\bibnamefont{Esteban}},
  \bibinfo{author}{\bibfnamefont{M.~C.} \bibnamefont{Gonzalez-Garcia}},
  \bibinfo{author}{\bibfnamefont{A.}~\bibnamefont{Hernandez-Cabezudo}},
  \bibinfo{author}{\bibfnamefont{M.}~\bibnamefont{Maltoni}}, \bibnamefont{and}
  \bibinfo{author}{\bibfnamefont{T.}~\bibnamefont{Schwetz}},
  \bibinfo{journal}{JHEP} \textbf{\bibinfo{volume}{01}}, \bibinfo{pages}{106}
  (\bibinfo{year}{2019}), \eprint{1811.05487}.

\bibitem[{\citenamefont{Tórtola et~al.}(2020)\citenamefont{Tórtola,
  Barenboim, and Ternes}}]{Tortola:2020ncu}
\bibinfo{author}{\bibfnamefont{M.~A.} \bibnamefont{Tórtola}},
  \bibinfo{author}{\bibfnamefont{G.~A.} \bibnamefont{Barenboim}},
  \bibnamefont{and} \bibinfo{author}{\bibfnamefont{C.~A.}
  \bibnamefont{Ternes}}, \bibinfo{journal}{JHEP} \textbf{\bibinfo{volume}{07}},
  \bibinfo{pages}{155} (\bibinfo{year}{2020}), \eprint{2005.05975}.

\bibitem[{\citenamefont{Kelly et~al.}(2020)\citenamefont{Kelly, Machado, Parke,
  Perez~Gonzalez, and Zukanovich-Funchal}}]{Kelly:2020fkv}
\bibinfo{author}{\bibfnamefont{K.~J.} \bibnamefont{Kelly}},
  \bibinfo{author}{\bibfnamefont{P.~A.} \bibnamefont{Machado}},
  \bibinfo{author}{\bibfnamefont{S.~J.} \bibnamefont{Parke}},
  \bibinfo{author}{\bibfnamefont{Y.~F.} \bibnamefont{Perez~Gonzalez}},
  \bibnamefont{and}
  \bibinfo{author}{\bibfnamefont{R.}~\bibnamefont{Zukanovich-Funchal}}
  (\bibinfo{year}{2020}), \eprint{2007.08526}.

\bibitem[{\citenamefont{Acciarri et~al.}(2015)}]{Acciarri:2015uup}
\bibinfo{author}{\bibfnamefont{R.}~\bibnamefont{Acciarri}} \bibnamefont{et~al.}
  (\bibinfo{collaboration}{DUNE}) (\bibinfo{year}{2015}), \eprint{1512.06148}.

\bibitem[{\citenamefont{Abi et~al.}(2020{\natexlab{a}})}]{Abi:2020evt}
\bibinfo{author}{\bibfnamefont{B.}~\bibnamefont{Abi}} \bibnamefont{et~al.}
  (\bibinfo{collaboration}{DUNE}) (\bibinfo{year}{2020}{\natexlab{a}}),
  \eprint{2002.03005}.

\bibitem[{\citenamefont{Abi et~al.}(2020{\natexlab{b}})}]{Abi:2020qib}
\bibinfo{author}{\bibfnamefont{B.}~\bibnamefont{Abi}} \bibnamefont{et~al.}
  (\bibinfo{collaboration}{DUNE}) (\bibinfo{year}{2020}{\natexlab{b}}),
  \eprint{2006.16043}.

\bibitem[{\citenamefont{Masud et~al.}(2019{\natexlab{a}})\citenamefont{Masud,
  Bishai, and Mehta}}]{Masud:2017bcf}
\bibinfo{author}{\bibfnamefont{M.}~\bibnamefont{Masud}},
  \bibinfo{author}{\bibfnamefont{M.}~\bibnamefont{Bishai}}, \bibnamefont{and}
  \bibinfo{author}{\bibfnamefont{P.}~\bibnamefont{Mehta}},
  \bibinfo{journal}{Sci. Rep.} \textbf{\bibinfo{volume}{9}},
  \bibinfo{pages}{352} (\bibinfo{year}{2019}{\natexlab{a}}),
  \eprint{1704.08650}.

\bibitem[{\citenamefont{Masud et~al.}(2019{\natexlab{b}})\citenamefont{Masud,
  Roy, and Mehta}}]{Masud:2018pig}
\bibinfo{author}{\bibfnamefont{M.}~\bibnamefont{Masud}},
  \bibinfo{author}{\bibfnamefont{S.}~\bibnamefont{Roy}}, \bibnamefont{and}
  \bibinfo{author}{\bibfnamefont{P.}~\bibnamefont{Mehta}},
  \bibinfo{journal}{Phys. Rev. D} \textbf{\bibinfo{volume}{99}},
  \bibinfo{pages}{115032} (\bibinfo{year}{2019}{\natexlab{b}}),
  \eprint{1812.10290}.

\bibitem[{\citenamefont{De~Gouvêa et~al.}(2019)\citenamefont{De~Gouvêa,
  Kelly, Stenico, and Pasquini}}]{deGouvea:2019ozk}
\bibinfo{author}{\bibfnamefont{A.}~\bibnamefont{De~Gouvêa}},
  \bibinfo{author}{\bibfnamefont{K.~J.} \bibnamefont{Kelly}},
  \bibinfo{author}{\bibfnamefont{G.}~\bibnamefont{Stenico}}, \bibnamefont{and}
  \bibinfo{author}{\bibfnamefont{P.}~\bibnamefont{Pasquini}},
  \bibinfo{journal}{Phys. Rev. D} \textbf{\bibinfo{volume}{100}},
  \bibinfo{pages}{016004} (\bibinfo{year}{2019}), \eprint{1904.07265}.

\bibitem[{\citenamefont{Ghoshal et~al.}(2019)\citenamefont{Ghoshal, Giarnetti,
  and Meloni}}]{Ghoshal:2019pab}
\bibinfo{author}{\bibfnamefont{A.}~\bibnamefont{Ghoshal}},
  \bibinfo{author}{\bibfnamefont{A.}~\bibnamefont{Giarnetti}},
  \bibnamefont{and} \bibinfo{author}{\bibfnamefont{D.}~\bibnamefont{Meloni}},
  \bibinfo{journal}{JHEP} \textbf{\bibinfo{volume}{12}}, \bibinfo{pages}{126}
  (\bibinfo{year}{2019}), \eprint{1906.06212}.

\bibitem[{\citenamefont{Huber et~al.}(2009)\citenamefont{Huber, Lindner,
  Schwetz, and Winter}}]{Huber:2009cw}
\bibinfo{author}{\bibfnamefont{P.}~\bibnamefont{Huber}},
  \bibinfo{author}{\bibfnamefont{M.}~\bibnamefont{Lindner}},
  \bibinfo{author}{\bibfnamefont{T.}~\bibnamefont{Schwetz}}, \bibnamefont{and}
  \bibinfo{author}{\bibfnamefont{W.}~\bibnamefont{Winter}},
  \bibinfo{journal}{JHEP} \textbf{\bibinfo{volume}{11}}, \bibinfo{pages}{044}
  (\bibinfo{year}{2009}), \eprint{0907.1896}.

\bibitem[{\citenamefont{Agarwalla et~al.}(2013)\citenamefont{Agarwalla,
  Prakash, and Sankar}}]{Agarwalla:2013ju}
\bibinfo{author}{\bibfnamefont{S.~K.} \bibnamefont{Agarwalla}},
  \bibinfo{author}{\bibfnamefont{S.}~\bibnamefont{Prakash}}, \bibnamefont{and}
  \bibinfo{author}{\bibfnamefont{S.}~\bibnamefont{Sankar}},
  \bibinfo{journal}{JHEP} \textbf{\bibinfo{volume}{07}}, \bibinfo{pages}{131}
  (\bibinfo{year}{2013}), \eprint{1301.2574}.

\bibitem[{\citenamefont{Machado et~al.}(2014)\citenamefont{Machado, Minakata,
  Nunokawa, and Zukanovich~Funchal}}]{Machado:2013kya}
\bibinfo{author}{\bibfnamefont{P.}~\bibnamefont{Machado}},
  \bibinfo{author}{\bibfnamefont{H.}~\bibnamefont{Minakata}},
  \bibinfo{author}{\bibfnamefont{H.}~\bibnamefont{Nunokawa}}, \bibnamefont{and}
  \bibinfo{author}{\bibfnamefont{R.}~\bibnamefont{Zukanovich~Funchal}},
  \bibinfo{journal}{JHEP} \textbf{\bibinfo{volume}{05}}, \bibinfo{pages}{109}
  (\bibinfo{year}{2014}), \eprint{1307.3248}.

\bibitem[{\citenamefont{Ghosh et~al.}(2017)\citenamefont{Ghosh, Goswami, and
  Raut}}]{Ghosh:2014zea}
\bibinfo{author}{\bibfnamefont{M.}~\bibnamefont{Ghosh}},
  \bibinfo{author}{\bibfnamefont{S.}~\bibnamefont{Goswami}}, \bibnamefont{and}
  \bibinfo{author}{\bibfnamefont{S.~K.} \bibnamefont{Raut}},
  \bibinfo{journal}{Mod. Phys. Lett. A} \textbf{\bibinfo{volume}{32}},
  \bibinfo{pages}{1750034} (\bibinfo{year}{2017}), \eprint{1409.5046}.

\bibitem[{\citenamefont{Ghosh et~al.}(2016{\natexlab{a}})\citenamefont{Ghosh,
  Ghoshal, Goswami, Nath, and Raut}}]{Ghosh:2015ena}
\bibinfo{author}{\bibfnamefont{M.}~\bibnamefont{Ghosh}},
  \bibinfo{author}{\bibfnamefont{P.}~\bibnamefont{Ghoshal}},
  \bibinfo{author}{\bibfnamefont{S.}~\bibnamefont{Goswami}},
  \bibinfo{author}{\bibfnamefont{N.}~\bibnamefont{Nath}}, \bibnamefont{and}
  \bibinfo{author}{\bibfnamefont{S.~K.} \bibnamefont{Raut}},
  \bibinfo{journal}{Phys. Rev. D} \textbf{\bibinfo{volume}{93}},
  \bibinfo{pages}{013013} (\bibinfo{year}{2016}{\natexlab{a}}),
  \eprint{1504.06283}.

\bibitem[{\citenamefont{Ghosh et~al.}(2016{\natexlab{b}})\citenamefont{Ghosh,
  Goswami, and Raut}}]{Ghosh:2014rna}
\bibinfo{author}{\bibfnamefont{M.}~\bibnamefont{Ghosh}},
  \bibinfo{author}{\bibfnamefont{S.}~\bibnamefont{Goswami}}, \bibnamefont{and}
  \bibinfo{author}{\bibfnamefont{S.~K.} \bibnamefont{Raut}},
  \bibinfo{journal}{Eur. Phys. J. C} \textbf{\bibinfo{volume}{76}},
  \bibinfo{pages}{114} (\bibinfo{year}{2016}{\natexlab{b}}),
  \eprint{1412.1744}.

\bibitem[{\citenamefont{Nath et~al.}(2016)\citenamefont{Nath, Ghosh, and
  Goswami}}]{Nath:2015kjg}
\bibinfo{author}{\bibfnamefont{N.}~\bibnamefont{Nath}},
  \bibinfo{author}{\bibfnamefont{M.}~\bibnamefont{Ghosh}}, \bibnamefont{and}
  \bibinfo{author}{\bibfnamefont{S.}~\bibnamefont{Goswami}},
  \bibinfo{journal}{Nucl. Phys. B} \textbf{\bibinfo{volume}{913}},
  \bibinfo{pages}{381} (\bibinfo{year}{2016}), \eprint{1511.07496}.

\bibitem[{\citenamefont{Coloma et~al.}(2014)\citenamefont{Coloma, Minakata, and
  Parke}}]{Coloma:2014kca}
\bibinfo{author}{\bibfnamefont{P.}~\bibnamefont{Coloma}},
  \bibinfo{author}{\bibfnamefont{H.}~\bibnamefont{Minakata}}, \bibnamefont{and}
  \bibinfo{author}{\bibfnamefont{S.~J.} \bibnamefont{Parke}},
  \bibinfo{journal}{Phys. Rev. D} \textbf{\bibinfo{volume}{90}},
  \bibinfo{pages}{093003} (\bibinfo{year}{2014}), \eprint{1406.2551}.

\bibitem[{\citenamefont{Adams et~al.}(2013)}]{Adams:2013qkq}
\bibinfo{author}{\bibfnamefont{C.}~\bibnamefont{Adams}} \bibnamefont{et~al.}
  (\bibinfo{collaboration}{LBNE}), in \emph{\bibinfo{booktitle}{{Snowmass
  2013}: {Workshop on Energy Frontier}}} (\bibinfo{year}{2013}),
  \eprint{1307.7335}.

\bibitem[{\citenamefont{Ballett et~al.}(2017)\citenamefont{Ballett, King,
  Pascoli, Prouse, and Wang}}]{Ballett:2016daj}
\bibinfo{author}{\bibfnamefont{P.}~\bibnamefont{Ballett}},
  \bibinfo{author}{\bibfnamefont{S.~F.} \bibnamefont{King}},
  \bibinfo{author}{\bibfnamefont{S.}~\bibnamefont{Pascoli}},
  \bibinfo{author}{\bibfnamefont{N.~W.} \bibnamefont{Prouse}},
  \bibnamefont{and} \bibinfo{author}{\bibfnamefont{T.}~\bibnamefont{Wang}},
  \bibinfo{journal}{Phys. Rev. D} \textbf{\bibinfo{volume}{96}},
  \bibinfo{pages}{033003} (\bibinfo{year}{2017}), \eprint{1612.07275}.

\bibitem[{bnl()}]{bnl_le}
\bibinfo{note}{DUNE Internal Document DUNE-doc-20412-v1}.

\bibitem[{\citenamefont{Huber et~al.}(2005)\citenamefont{Huber, Lindner, and
  Winter}}]{Huber:2004ka}
\bibinfo{author}{\bibfnamefont{P.}~\bibnamefont{Huber}},
  \bibinfo{author}{\bibfnamefont{M.}~\bibnamefont{Lindner}}, \bibnamefont{and}
  \bibinfo{author}{\bibfnamefont{W.}~\bibnamefont{Winter}},
  \bibinfo{journal}{Comput. Phys. Commun.} \textbf{\bibinfo{volume}{167}},
  \bibinfo{pages}{195} (\bibinfo{year}{2005}), \eprint{hep-ph/0407333}.

\bibitem[{\citenamefont{Huber et~al.}(2007)\citenamefont{Huber, Kopp, Lindner,
  Rolinec, and Winter}}]{Huber:2007ji}
\bibinfo{author}{\bibfnamefont{P.}~\bibnamefont{Huber}},
  \bibinfo{author}{\bibfnamefont{J.}~\bibnamefont{Kopp}},
  \bibinfo{author}{\bibfnamefont{M.}~\bibnamefont{Lindner}},
  \bibinfo{author}{\bibfnamefont{M.}~\bibnamefont{Rolinec}}, \bibnamefont{and}
  \bibinfo{author}{\bibfnamefont{W.}~\bibnamefont{Winter}},
  \bibinfo{journal}{Comput. Phys. Commun.} \textbf{\bibinfo{volume}{177}},
  \bibinfo{pages}{432} (\bibinfo{year}{2007}), \eprint{hep-ph/0701187}.

\bibitem[{\citenamefont{Dziewonski and Anderson}(1981)}]{Dziewonski:1981xy}
\bibinfo{author}{\bibfnamefont{A.~M.} \bibnamefont{Dziewonski}}
  \bibnamefont{and} \bibinfo{author}{\bibfnamefont{D.~L.}
  \bibnamefont{Anderson}}, \bibinfo{journal}{Phys. Earth Planet. Interiors}
  \textbf{\bibinfo{volume}{25}}, \bibinfo{pages}{297} (\bibinfo{year}{1981}).

\bibitem[{\citenamefont{Gandhi et~al.}(2005)\citenamefont{Gandhi, Ghoshal,
  Goswami, Mehta, and Sankar}}]{Gandhi:2004md}
\bibinfo{author}{\bibfnamefont{R.}~\bibnamefont{Gandhi}},
  \bibinfo{author}{\bibfnamefont{P.}~\bibnamefont{Ghoshal}},
  \bibinfo{author}{\bibfnamefont{S.}~\bibnamefont{Goswami}},
  \bibinfo{author}{\bibfnamefont{P.}~\bibnamefont{Mehta}}, \bibnamefont{and}
  \bibinfo{author}{\bibfnamefont{S.~U.} \bibnamefont{Sankar}},
  \bibinfo{journal}{Phys.Rev.Lett.} \textbf{\bibinfo{volume}{94}},
  \bibinfo{pages}{051801} (\bibinfo{year}{2005}), \eprint{hep-ph/0408361}.

\bibitem[{\citenamefont{Gandhi et~al.}(2006)\citenamefont{Gandhi, Ghoshal,
  Goswami, Mehta, and Sankar}}]{Gandhi:2004bj}
\bibinfo{author}{\bibfnamefont{R.}~\bibnamefont{Gandhi}},
  \bibinfo{author}{\bibfnamefont{P.}~\bibnamefont{Ghoshal}},
  \bibinfo{author}{\bibfnamefont{S.}~\bibnamefont{Goswami}},
  \bibinfo{author}{\bibfnamefont{P.}~\bibnamefont{Mehta}}, \bibnamefont{and}
  \bibinfo{author}{\bibfnamefont{S.~U.} \bibnamefont{Sankar}},
  \bibinfo{journal}{Phys.Rev.} \textbf{\bibinfo{volume}{D73}},
  \bibinfo{pages}{053001} (\bibinfo{year}{2006}), \eprint{hep-ph/0411252}.

\bibitem[{\citenamefont{Kelly and Parke}(2018)}]{Kelly:2018kmb}
\bibinfo{author}{\bibfnamefont{K.~J.} \bibnamefont{Kelly}} \bibnamefont{and}
  \bibinfo{author}{\bibfnamefont{S.~J.} \bibnamefont{Parke}},
  \bibinfo{journal}{Phys. Rev. D} \textbf{\bibinfo{volume}{98}},
  \bibinfo{pages}{015025} (\bibinfo{year}{2018}), \eprint{1802.06784}.

\bibitem[{\citenamefont{Chatterjee et~al.}(2018)\citenamefont{Chatterjee,
  Kamiya, Moura, and Yu}}]{Chatterjee:2018dyd}
\bibinfo{author}{\bibfnamefont{A.}~\bibnamefont{Chatterjee}},
  \bibinfo{author}{\bibfnamefont{F.}~\bibnamefont{Kamiya}},
  \bibinfo{author}{\bibfnamefont{C.~A.} \bibnamefont{Moura}}, \bibnamefont{and}
  \bibinfo{author}{\bibfnamefont{J.}~\bibnamefont{Yu}} (\bibinfo{year}{2018}),
  \eprint{1809.09313}.

\bibitem[{\citenamefont{Alion et~al.}(2016)}]{Alion:2016uaj}
\bibinfo{author}{\bibfnamefont{T.}~\bibnamefont{Alion}} \bibnamefont{et~al.}
  (\bibinfo{collaboration}{DUNE}) (\bibinfo{year}{2016}), \eprint{1606.09550}.

\bibitem[{\citenamefont{Agostinelli et~al.}(2003)}]{Agostinelli:2002hh}
\bibinfo{author}{\bibfnamefont{S.}~\bibnamefont{Agostinelli}}
  \bibnamefont{et~al.} (\bibinfo{collaboration}{GEANT4}),
  \bibinfo{journal}{Nucl. Instrum. Meth.} \textbf{\bibinfo{volume}{A506}},
  \bibinfo{pages}{250} (\bibinfo{year}{2003}).

\bibitem[{\citenamefont{Allison et~al.}(2006)}]{Allison:2006ve}
\bibinfo{author}{\bibfnamefont{J.}~\bibnamefont{Allison}} \bibnamefont{et~al.},
  \bibinfo{journal}{IEEE Trans. Nucl. Sci.} \textbf{\bibinfo{volume}{53}},
  \bibinfo{pages}{270} (\bibinfo{year}{2006}).

\bibitem[{\citenamefont{Huber et~al.}(2002)\citenamefont{Huber, Lindner, and
  Winter}}]{Huber:2002mx}
\bibinfo{author}{\bibfnamefont{P.}~\bibnamefont{Huber}},
  \bibinfo{author}{\bibfnamefont{M.}~\bibnamefont{Lindner}}, \bibnamefont{and}
  \bibinfo{author}{\bibfnamefont{W.}~\bibnamefont{Winter}},
  \bibinfo{journal}{Nucl. Phys.} \textbf{\bibinfo{volume}{B645}},
  \bibinfo{pages}{3} (\bibinfo{year}{2002}), \eprint{hep-ph/0204352}.

\bibitem[{\citenamefont{Fogli et~al.}(2002)\citenamefont{Fogli, Lisi, Marrone,
  Montanino, and Palazzo}}]{Fogli:2002pt}
\bibinfo{author}{\bibfnamefont{G.~L.} \bibnamefont{Fogli}},
  \bibinfo{author}{\bibfnamefont{E.}~\bibnamefont{Lisi}},
  \bibinfo{author}{\bibfnamefont{A.}~\bibnamefont{Marrone}},
  \bibinfo{author}{\bibfnamefont{D.}~\bibnamefont{Montanino}},
  \bibnamefont{and} \bibinfo{author}{\bibfnamefont{A.}~\bibnamefont{Palazzo}},
  \bibinfo{journal}{Phys. Rev.} \textbf{\bibinfo{volume}{D66}},
  \bibinfo{pages}{053010} (\bibinfo{year}{2002}), \eprint{hep-ph/0206162}.

\bibitem[{\citenamefont{Gonzalez-Garcia and
  Maltoni}(2004)}]{GonzalezGarcia:2004wg}
\bibinfo{author}{\bibfnamefont{M.}~\bibnamefont{Gonzalez-Garcia}}
  \bibnamefont{and} \bibinfo{author}{\bibfnamefont{M.}~\bibnamefont{Maltoni}},
  \bibinfo{journal}{Phys.Rev.} \textbf{\bibinfo{volume}{D70}},
  \bibinfo{pages}{033010} (\bibinfo{year}{2004}), \eprint{hep-ph/0404085}.

\bibitem[{\citenamefont{Gandhi et~al.}(2007)\citenamefont{Gandhi, Ghoshal,
  Goswami, Mehta, Sankar, and Shalgar}}]{Gandhi:2007td}
\bibinfo{author}{\bibfnamefont{R.}~\bibnamefont{Gandhi}},
  \bibinfo{author}{\bibfnamefont{P.}~\bibnamefont{Ghoshal}},
  \bibinfo{author}{\bibfnamefont{S.}~\bibnamefont{Goswami}},
  \bibinfo{author}{\bibfnamefont{P.}~\bibnamefont{Mehta}},
  \bibinfo{author}{\bibfnamefont{S.~U.} \bibnamefont{Sankar}},
  \bibnamefont{and} \bibinfo{author}{\bibfnamefont{S.}~\bibnamefont{Shalgar}},
  \bibinfo{journal}{Phys. Rev.} \textbf{\bibinfo{volume}{D76}},
  \bibinfo{pages}{073012} (\bibinfo{year}{2007}), \eprint{0707.1723}.

\bibitem[{\citenamefont{Qian et~al.}(2012)\citenamefont{Qian, Tan, Wang, Ling,
  McKeown, and Zhang}}]{Qian:2012zn}
\bibinfo{author}{\bibfnamefont{X.}~\bibnamefont{Qian}},
  \bibinfo{author}{\bibfnamefont{A.}~\bibnamefont{Tan}},
  \bibinfo{author}{\bibfnamefont{W.}~\bibnamefont{Wang}},
  \bibinfo{author}{\bibfnamefont{J.~J.} \bibnamefont{Ling}},
  \bibinfo{author}{\bibfnamefont{R.~D.} \bibnamefont{McKeown}},
  \bibnamefont{and} \bibinfo{author}{\bibfnamefont{C.}~\bibnamefont{Zhang}},
  \bibinfo{journal}{Phys. Rev.} \textbf{\bibinfo{volume}{D86}},
  \bibinfo{pages}{113011} (\bibinfo{year}{2012}), \eprint{1210.3651}.

\bibitem[{\citenamefont{Masud et~al.}(2016)\citenamefont{Masud, Chatterjee, and
  Mehta}}]{Masud:2015xva}
\bibinfo{author}{\bibfnamefont{M.}~\bibnamefont{Masud}},
  \bibinfo{author}{\bibfnamefont{A.}~\bibnamefont{Chatterjee}},
  \bibnamefont{and} \bibinfo{author}{\bibfnamefont{P.}~\bibnamefont{Mehta}},
  \bibinfo{journal}{J. Phys.} \textbf{\bibinfo{volume}{G43}},
  \bibinfo{pages}{095005} (\bibinfo{year}{2016}), \eprint{1510.08261}.

\bibitem[{\citenamefont{Masud and Mehta}(2016{\natexlab{a}})}]{Masud:2016bvp}
\bibinfo{author}{\bibfnamefont{M.}~\bibnamefont{Masud}} \bibnamefont{and}
  \bibinfo{author}{\bibfnamefont{P.}~\bibnamefont{Mehta}},
  \bibinfo{journal}{Phys. Rev.} \textbf{\bibinfo{volume}{D94}},
  \bibinfo{pages}{013014} (\bibinfo{year}{2016}{\natexlab{a}}),
  \eprint{1603.01380}.

\bibitem[{\citenamefont{Masud and Mehta}(2016{\natexlab{b}})}]{Masud:2016gcl}
\bibinfo{author}{\bibfnamefont{M.}~\bibnamefont{Masud}} \bibnamefont{and}
  \bibinfo{author}{\bibfnamefont{P.}~\bibnamefont{Mehta}},
  \bibinfo{journal}{Phys. Rev.} \textbf{\bibinfo{volume}{D94}},
  \bibinfo{pages}{053007} (\bibinfo{year}{2016}{\natexlab{b}}),
  \eprint{1606.05662}.

\end{thebibliography}

\end{document}